\documentclass[aps,showpacs,preprintnumbers,amsmath,amssymb]{revtex4}
\usepackage{dcolumn}
\usepackage[dvips]{epsfig}

\begin{document}
\baselineskip=0.8 cm
\title{\bf Dynamical evolution of a vector field perturbation coupling to Einstein tensor}

\author{Songbai Chen\footnote{csb3752@hunnu.edu.cn}, Jiliang Jing
\footnote{jljing@hunnu.edu.cn}}

\affiliation{Institute of Physics and Department of Physics, Hunan
Normal University,  Changsha, Hunan 410081, People's Republic of
China \\ Key Laboratory of Low Dimensional Quantum Structures \\
and Quantum Control of Ministry of Education, Hunan Normal
University, Changsha, Hunan 410081, People's Republic of China\\
Kavli Institute for Theoretical Physics China, CAS, Beijing 100190,China}

\begin{abstract}
\baselineskip=0.6 cm
\begin{center}
{\bf Abstract}
\end{center}

We have investigated the wave dynamics of a vector field perturbation coupling to Einstein tensor in the four-dimensional Reissner-Nordstr\"{o}m black hole spacetime. Our results show that besides the dependence on the coupling between the vector field and Einstein tensor, the wave dynamic equation of the vector field perturbation strongly depends on the parity of the perturbation itself, which is quite different from that of the usual vector field perturbation without the coupling in the four-dimensional spacetime. Moreover, we also find that the vector field perturbation with odd parity grows with exponential rate if the coupling strength is stronger than certain a critical value. However, the vector field perturbation with even parity always decays in the Reissner-Nordstr\"{o}m black hole spacetime.

\end{abstract}

 \maketitle
\newpage
\section{Introduction}

In the Standard Model of particle physics, vector field associated with spin-1 describes gauge bosons including the photon, gluon, W and Z bosons, which means that vector field plays the important roles in the fundamental
physical theory. Electromagnetic field is a special kind of vector fields to describe the electromagnetic force between charged elementary particles mediated by gauge bosons. It is well known that the electromagnetic force and gravity are two kinds of fundamental forces in nature, which implies that the coupling between the electromagnetic and gravitational fields should be a kind of important interactions in physics. In the standard Einstein-Maxwell theory, the Lagrangian contains only the Einstein-Hilbert term and the quadratic term of Maxwell tensor. The latter is related directly to electromagnetic field, which can also be regarded as the coupling between Maxwell tensor and the spacetime metric tensor. But the interactions between electromagnetic field and curvature tensor are excluded in this standard electromagnetic theory.
In general, such a kind of couplings related to spacetime curvature in Lagrangian could give rise to higher order terms both in the Maxwell and Einstein equations, which results in the more complicated and strange behaviors of the electromagnetic field in the background spacetime.
Recently, a special kind of generalized Einstein-Maxwell theories of electromagnetic field coupling to curvature tensor have been investigated in \cite{Balakin,Faraoni,Hehl,Balakin1}. It is shown that in these models the equations of motion both for the electromagnetic and gravitational fields are still second-order differential equations, and
the coupling terms modify only the coefficients of the second-order derivatives. However, these modifications change the properties of gravitational and electromagnetic waves propagated in the spacetime and  yield time delays in the arrival of these waves \cite{Balakin}. The electromagnetic quantum fluctuations caused by these couplings can also lead to the inflation in the evolution of the early Universe\cite{Turner,Mazzitelli,Lambiase,Raya,Campanelli}. Morever, it is found that these fluctuations could be used as a mechanism to interpret
the large scale magnetic fields observed in clusters of galaxies \cite{Bamba,Kim,Clarke}.

Weyl tensor is an important tensor in general relativity, which describes a type of gravitational distortion in the spacetime. The couplings between Maxwell field and Weyl tensor have been investigated extensively in the literature \cite{Weyl1,Drummond,Dereli1,Solanki,Wu2011,Ma2011,Momeni,Roychowdhury,zhao2013}. Considering that Weyl tensor is related to the curvature tensors $R_{\mu\nu\rho\sigma}$, $R_{\mu\nu}$ and the Ricci scalar $R$,  the electrodynamics coupling to Weyl tensor can be understand as a special kind of generalized Einstein-Maxwell theory with the couplings between electromagnetic field and curvature tensors.
The coupling terms with Weyl tensor could be emerged naturally in quantum electrodynamics with the photon effective action arising from one-loop vacuum polarization on a curved background spacetime \cite{Drummond}.
Moreover, the recent investigations also imply that these couplings should be appeared in the strong gravitational region of classical compact astrophysical objects with high mass density at the center of galaxies \cite{Dereli1,Solanki}. The effects of the coupling between Maxwell field and Weyl tensor on holographic conductivity and charge diffusion in the anti-de Sitter spacetime are studied in \cite{Weyl1}, which is shown that the presence of the coupling changes the universal
relation with the $U(1)$ central charge observed at leading order.
Moreover,  it is found that the couplings with Weyl tensor modify the critical temperature at which holographic superconductors happen and change the order of the phase transition of the holographic superconductor \cite{Wu2011,Ma2011,Momeni,Roychowdhury}. The effects of such a kind of couplings on the transition between the holographic insulator phase and
superconductor phase have been also investigated in the AdS soliton spacetime \cite{zhao2013}. Recently, we studied the dynamical evolution and Hawking radiation of the electromagnetic field coupling to Weyl tensor in the Schwarzschild black hole spacetime \cite{sb2013,sb2014}. Our results show that both the dynamical evolution and Hawking radiation of the electromagnetic field depend not only on the coupling parameter, but also on the parity of the electromagnetic field.

The couplings between scalar field and curvature tensor has been also investigated in \cite{Sushkov,Gao,Granda,Sarida,James,Kaix}.
The investigation indicates \cite{Sushkov} that in cosmology the coupling between Einstein tensor and scalar field $G^{\mu\nu}\partial_{\mu}\psi\partial_{\nu}\psi$ can solve naturally
the problem of a graceful exit from inflation without
any fine-tuned potential. Moreover, the cosmic evolution of a scalar field
with the kinetic term coupling to more than one Einstein tensor indicates that
the scalar field behaves nearly like a dynamic cosmological constant during the late time evolution of the Universe \cite{Gao}. The effects of the kinetic couplings with Einstein tensor in the cosmology have also been investigated in Refs. \cite{Granda,Sarida,James,Kaix}. We \cite{sb2010} studied the Hawking radiation and dynamical evolution for a scalar field
coupling to Einstein¡¯s tensor in the background of the
Reissner-Nordstr\"{o}m black hole spacetime. It is shown that
the couplings between scalar field and Einstein's tensor not only enhances the absorption
probability and Hawking radiation of the black hole, and but also modifies the standard results of dynamical evolution of the scalar field in the background spacetime. The scalar hairs from a derivative coupling of a scalar field to the Einstein tensor have been investigated in Braneworlds \cite{Mas} and the scalar-tensor theory \cite{Mas1}. With this coupling, the study also shows \cite{Theod} that a phase transition of a Reissner-Nordstr\"{o}m black hole to a hairy black hole occurs in asymptotically flat spacetime since an Abelian $U(1)$ gauge symmetry was broken in the vicinity of the horizon of the black hole as the coupling constant is large enough. These new features of the coupled matter fields
may attract more attention to
study the couplings among matter fields and spacetime curvature tensors in the more general cases.

In order to probe the universal features of the matter field coupling to spacetime curvature tensors, we here will study the dynamical evolution of an vector field perturbation coupling to Einstein tensor in the background spacetime, and then explore the effect of the coupling on the quasinormal modes and the stability of the vector field perturbation in the black hole spacetime. The plan of our paper is organized as follows: in Sec.II, we will construct a simple form of the coupling between vector field and Einstein tensor and then derive the master equation of the coupled vector field perturbation in the four-dimensional static and spherical
symmetric black hole spacetime. In Sec.III, we will
study numerically the effects of the coupling on the quasinormal modes of the
external vector field perturbation in the Reissner-Nordstr\"{o}m black hole spacetime and then examine if
the perturbation is stable or not in this case. We also make a comparison among the dynamical properties of perturbational fields coupling to gravitational tensors in some black hole spacetimes.  To conclude, we present a summary in the last section.

\section{Wave equation for the vector field perturbations coupling to Einstein tensor}

Theoretically, there exists a lot of coupling forms between the vector field and Einstein tensor. The simplest form of the Lagrangian density with this coupling in the curved spacetime can be expressed as
\begin{eqnarray}
L_{EM}=-\frac{1}{4}\bigg(F_{\mu\nu}F^{\mu\nu}-4\alpha
G^{\mu\rho}F_{\mu\nu}F_{\rho}^{\;\nu}\bigg),\label{acts}
\end{eqnarray}
where $G_{\mu\nu}$ is the Einstein tensor and $\alpha$ is a
coupling constant with dimensions of length-squared. $F_{\mu\nu}$ is
the Maxwell tensor, which is related to the four
vector potential $A_{\mu}$ by $F_{\mu\nu}=A_{\nu;\mu}-A_{\mu;\nu}$.
As in the case with Weyl corrections \cite{Weyl1,Drummond,Dereli1,Solanki,Wu2011,Ma2011,Momeni,Roychowdhury,zhao2013}, the equation of motion of the vector field is corrected by the coupling term with Einstein tensor, which is a tensorial structure modifying the equation at leading order in derivatives.
One of the advantages of the coupling form we chosen here is that the modified master equation for the vector field perturbation can be
decoupled to a second order differential equation. It is very crucial
for us to study further the dynamical evolution of
the perturbation in the black hole spacetime.

Varying the action (\ref{acts}) with respect to $A_{\mu}$,  it is easy to
get the equations of motion of the vector field coupling to Einstein tensor
\begin{eqnarray}
\nabla_{\mu}\bigg(F^{\mu\nu}-2\alpha
G^{\mu\rho}F_{\rho}^{\;\nu}+2\alpha
G^{\nu\rho}F_{\rho}^{\;\mu}\bigg)=0.\label{WE}
\end{eqnarray}
Obviously, the presence of the coupling terms $G^{\mu\rho}F_{\rho}^{\;\nu}$ and $G^{\nu\rho}F_{\rho}^{\;\mu}$  in the  equation of motion implies that the coupling between the vector field and Einstein tensor will modify the dynamical behaviors of
the vector field perturbation in the background spacetime.

In the background of
a four-dimensional static and spherical
symmetric black hole spacetime with the metric
\begin{eqnarray}
ds^2&=&fdt^2-\frac{1}{f}dr^2-r^2
d\theta^2-r^2\sin^2{\theta}d\phi^2,\label{m1}
\end{eqnarray}
we can expand the four vector potential $A_{\mu}$ in vector spherical harmonics \cite{Ruffini,Chandrasekhar1983}
\begin{eqnarray}
A_{\mu}= \sum_{l,m}\left(\left[\begin{array}{ccc}
 &0&\\
 &0&\\
 &\frac{a^{lm}(t,r)}{\sin\theta}\partial_{\phi}Y_{lm}&\\
 &-a^{lm}(t,r)\sin\theta\partial_{\theta}Y_{lm}&
\end{array}\right]+\left[\begin{array}{cccc}
 &j^{lm}(t,r)Y_{lm}&\\
 &h^{lm}(t,r)Y_{lm}&\\
 &k^{lm}(t,r)\partial_{\theta}Y_{lm}&\\
 &k^{lm}(t,r)\partial_{\phi}Y_{lm}&
\end{array}\right]\right),\label{Au}
\end{eqnarray}
where the metric coefficient $f$ is a function of polar coordinate $r$.
The first term in the vector potential (\ref{Au}) has parity $(-1)^{l+1}$ and
the second term has parity $(-1)^{l}$, where $l$ is the angular quantum
number and $m$ is the azimuthal number. Since all of metric coefficients in the spacetime (\ref{m1}) are independent of time coordinate $t$, one can separate the variables $a^{lm}(t,r)$, $h^{lm}(t,r)$, $j^{lm}(t,r)$ and $k^{lm}(t,r)$ as
\begin{eqnarray}
a^{lm}(t,r)&=&a^{lm}(r)e^{-i\omega t},~~~~~h^{lm}(t,r)=h^{lm}(r)e^{-i\omega t},\nonumber\\
j^{lm}(t,r)&=&j^{lm}(r)e^{-i\omega t},~~~~~k^{lm}(t,r)=k^{lm}(r)e^{-i\omega t}.\label{Au1}
\end{eqnarray}
Substituting
the vector potential (\ref{Au}) with the separating variables (\ref{Au1}) into the modified equation of motion (\ref{WE}), one can obtain three independent coupled differential equations after some tedious calculations (See in appendix). Eliminating $k^{lm}(r)$, we find that both of the wave equations for the vector field perturbation with different parities can be simplified as the form of a
second order differential equation
\begin{eqnarray}
\frac{d^2\Psi(r)}{dr^2_*}+[\omega^2-V(r)]\Psi(r)=0,\label{radial}
\end{eqnarray}
where $r_{*}$ is the tortoise coordinate defined by $dr_*=\frac{
dr}{f}$. The wavefunction $\Psi(r)$ is a linear combination of the three functions appeared in the vector potential $A_{\mu}$, i.e., $j^{lm}(r)$, $h^{lm}(r)$, and $a^{lm}(r)$.  For the odd parity $(-1)^{l+1}$, the wavefunction $\Psi(r)$ can be expressed as
\begin{eqnarray}
\Psi(r)=a^{lm}\sqrt{1+\frac{\alpha}{r^2}(r^2f''+4rf'+2f-2)},
\end{eqnarray}
but for the even parity $(-1)^{l}$, it is given by
\begin{eqnarray}
\Psi(r)=\frac{r^2}{l(l+1)}\bigg(-i\omega
h^{lm}-\frac{dj^{lm}}{dr}\bigg)
\frac{\sqrt{1-\frac{\alpha}{r^2}(r^2f''+4rf'+2f-2)}}{1-\frac{4\alpha}{r^2}(rf'+f-1)}.
\end{eqnarray}
It is clear that the concrete form of $\Psi(r)$ depends not only on the coupling between the field and Einstein tensor, but also on the
parity of the vector field perturbation itself.
Correspondingly, the form of the potential $V(r)$  in Eq. (\ref{radial})  can be expressed as
\begin{eqnarray}
V(r)=f\bigg\{\frac{l(l+1)}{r^2}\bigg[\frac{1+\frac{2\alpha}{r}(rf''+2f')}{
1+\frac{\alpha}{r^2}(r^2f''+4rf'+2f-2)}\bigg]+\frac{\alpha(h_0+h_1\alpha)}{4r^2[r^2+
\alpha(r^2f''+4rf'+2f-2)]^2}\bigg\},\label{evod}
\end{eqnarray}
for the odd parity, and
\begin{eqnarray}
V(r)=f\bigg\{\frac{l(l+1)}{r^2}\bigg[\frac{1-\frac{\alpha}{r^2}(rf''+4f'+2f-2)}{
1-\frac{4\alpha}{r^2}(rf'+f-1)}\bigg]+\frac{\alpha(h_0+h_2\alpha)}{4r^2[r^2+
\alpha(r^2f''+4rf'+2f-2)]^2}\bigg\},\label{even}
\end{eqnarray}
for the even parity, respectively. Here the quantities $h_0$, $h_1$ and $h_2$ are
\begin{eqnarray}
h_0&=&2r^2[12f^2+rf'(r^3f^{(3)}+4r^2f''-2rf+4)
+f(r^4f^{(4)}+4r^3f^{(3)}-6r^2f''-4rf'-12)],
\\
h_1&=&2rf'(r^2 f''+4rf'-2)(r^3 f^{(3)}+4 r^2 f''-2 r f'+4)+4f^2(r^4 f^{(4)}+6 r^3 f^{(3)}+8 r^2 f''+16 rf' -16)\nonumber\\&+&f[2r^4f^{(4)}(r^2 f''+4 r f'-2)-r^3f^{(3)}(r^3f^{(3)}-40rf'+24)-4r^2f''(7r^2 f''+6rf'+8)]\nonumber\\&-&4f(11rf'^2+12rf'-8f^2-8),
\\
h_2&=&16r^3f'^3-4r^2f'^2(2r^3f^{(3)}+7r^2f''-13f+10)
+r^2f[60r^2f''^2-4r^2(f-1)f^{(4)}\nonumber\\&+&rf^{(3)}(3r^3f^{(3)}-40f+40) -2 (r^4 f^{(4)}-8 r^3 f^{(3)}+48 f-48) f'']\nonumber\\&-&2rf'[(r^2 f''-2)(r^3 f^{(3)}+4r^2 f''+4)+4 r^2f(r^2 f^{(4)}+6 r f^{(3)}+f'')+8f(2f-1)]
.
\end{eqnarray}
It is easy to find that the effective potentials $V(r)$ depend on the parities of the vector field perturbations, which
is different from that in the usual case without the coupling between vector field and Einstein tensor in which the effective potential of the vector field perturbation is independent of the parity of field.  The similar behaviors of the effective potential is also found in the case of the electromagnetic perturbations with Weyl corrections in the four dimensional spacetime \cite{sb2013}. It could be explained by a fact that the coupling terms $C_{\mu\nu\rho\sigma}F^{\mu\nu}F^{\rho\sigma}$ and $G^{\mu\rho}F_{\mu\nu}F_{\rho}^{\;\nu}$ can be treated as a kind of general classical couplings between the gravitational tensor and the matter field since both of the Weyl tensor $C_{\mu\nu\rho\sigma}$ and the Einstein tensor $G_{\mu\nu}$ are functions of the curvature tensors of background spacetime. Moreover, we also find that the effective potentials (\ref{evod}) and (\ref{even}) can be recovered back to the usual form for the vector field perturbation without coupling to Einstein tensor as the coupling constant $\alpha=0$, in which the effective potentials for different parities share the same form.

\section{Wave dynamics of the vector field perturbation coupling to Einstein tensor in the Reissner-Nordstr\"{o}m black hole spacetime}

We are now in position to study the wave dynamics of the vector field perturbation coupling to Einstein tensor in a four dimensional Reissner-Nordstr\"{o}m black hole spacetime. Here, we choose the Reissner-Nordstr\"{o}m black hole spacetime as a background metric mainly because it is the simplest black hole with the nonzero
components of the Einstein tensor $G_{\mu\nu}$ in general
relativity theory. In this system, the action can be expressed as
\begin{eqnarray}
S=\int d^4x \sqrt{-g}\bigg[\frac{R}{16\pi
G}-\frac{1}{4}\mathcal{F}_{\mu\nu}\mathcal{F}^{\mu\nu}-\frac{1}{4}\bigg(F_{\mu\nu}F^{\mu\nu}-4\alpha
G^{\mu\rho}F_{\mu\nu}F_{\rho}^{\;\nu}\bigg)\bigg].\label{acts1}
\end{eqnarray}
Obviously, there exist two different kinds of matter fields. One of them is the free electromagnetic field described by the four-vector potential $\mathcal{A}_{\mu}$ and its electromagnetic field strength tensor
is $\mathcal{F}_{\mu\nu}=\partial_{\mu}\mathcal{A}_{\nu}-
\partial_{\nu}\mathcal{A}_{\mu}$, which satisfies the usual Maxwell equations $\nabla_{\mu}\mathcal{F}^{\mu\nu}=0$. The other is the vector field $A_{\mu}$ interacting with Einstein tensor, which obeys the modified equations of motion (\ref{WE}). This means that the equations of motion for the coupled vector field $A_{\mu}$ is different from that for the free electromagnetic field $\mathcal{A}_{\mu}$, which is understandable because the electromagnetic field is only a special kind of vector fields. The systems with the electromagnetic field and the other vector field have been investigated extensively in \cite{Wangm1,Ferreira,Witten,RG1}. Here, we treat the vector field  $A_{\mu}$ interacting with Einstein tensor as an external perturbational field, which doesn't affect the metric of the background spacetime because the back-reaction of the perturbational field can be neglected. Therefore,
Reissner-Nordstr\"{o}m black hole is an exact solution of the system (\ref{acts1}) unperturbed by the external vector field $A_{\mu}$, and we can study the dynamical behaviors of the vector field perturbation $A_{\mu}$ in this background.

For a four dimensional Reissner-Nordstr\"{o}m black hole spacetime, the metric function is
$f=1-\frac{2M}{r}+\frac{q^2}{r^2}$. Substituting it into Eqs.(\ref{evod}) and (\ref{even}), one can obtain the effective potentials for the vector field perturbations
\begin{eqnarray}
V(r)_{odd}&=&\bigg(1-\frac{2M}{r}+\frac{q^2}{r^2}\bigg)\frac{l(l+1)}{r^2}
\frac{r^4+4\alpha
q^2}{r^4},\nonumber\\ \label{evodsch}
\end{eqnarray}
for the odd parity and
\begin{eqnarray}
V(r)_{even}=\bigg(1-\frac{2M}{r}+\frac{q^2}{r^2}\bigg)\frac{l(l+1)r^2}{r^4+4\alpha q^2}
,\label{evensch}
\end{eqnarray}
for the even parity, respectively. From above formulas, it is easy to find that the effective potential for odd parity perturbation (\ref{evodsch}) is
continuous in the region outside the black hole, but the effective potential for even parity perturbation (\ref{evensch}) is discontinuous and divergent at the point $r^4+4\alpha q^2=0$. Considering that the effective potential should be continuous in the physical region outside the black hole event, we here must impose a constraint on the coupling constant $\alpha$ for the vector field perturbations with even purturbation, i.e., $r^4_++4\alpha q^2>0$,  which ensures that the effective potential (\ref{evensch}) is continuous outside the outer event horizon of Reissner-Nordstr\"{o}m black hole.  The constraint depends on the parities of vector fields, which is similar to that in the case of Weyl corrections \cite{sb2013}.

\begin{figure}[ht]
\begin{center}
\includegraphics[width=5.5cm]{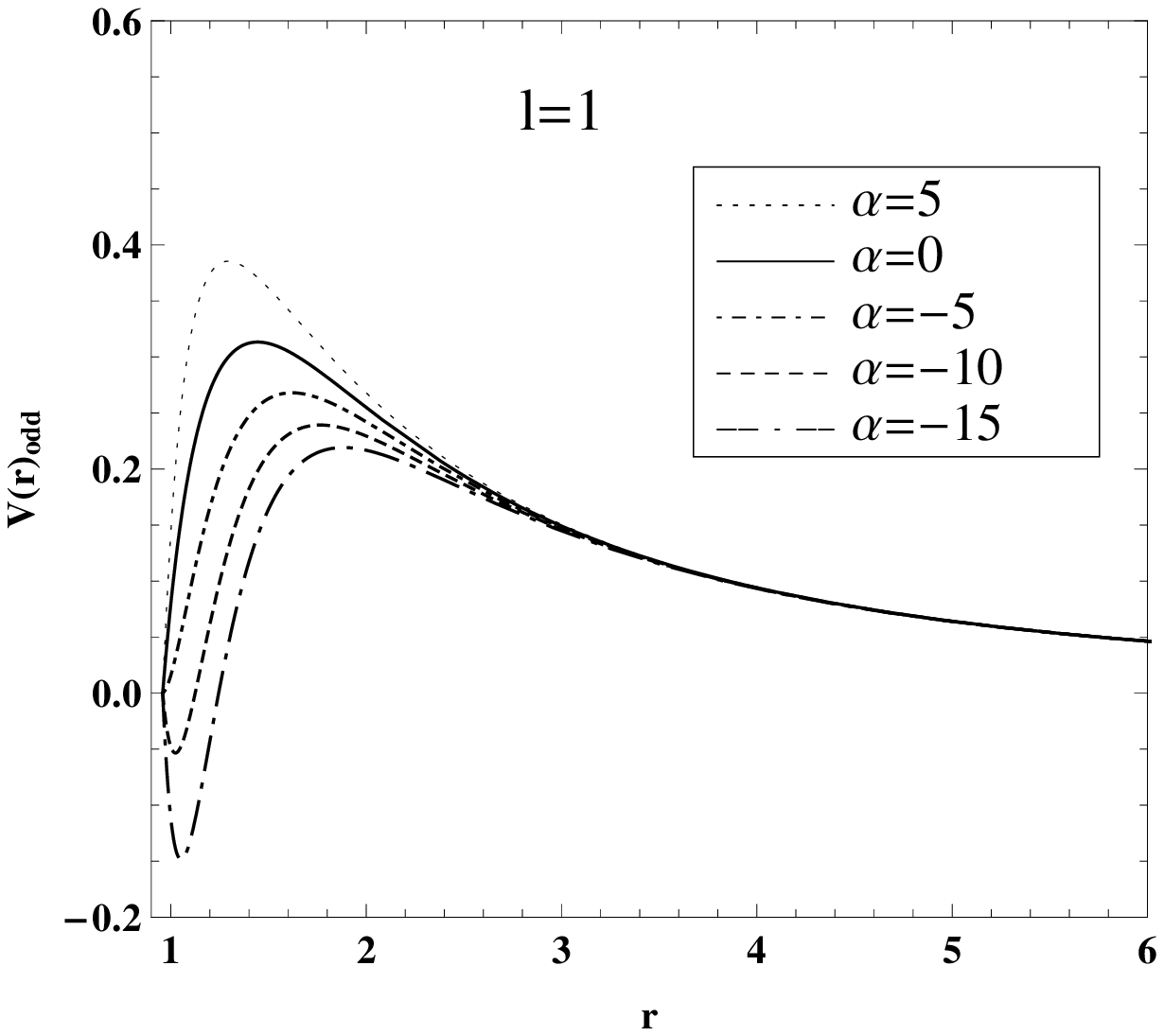}
\includegraphics[width=5.5cm]{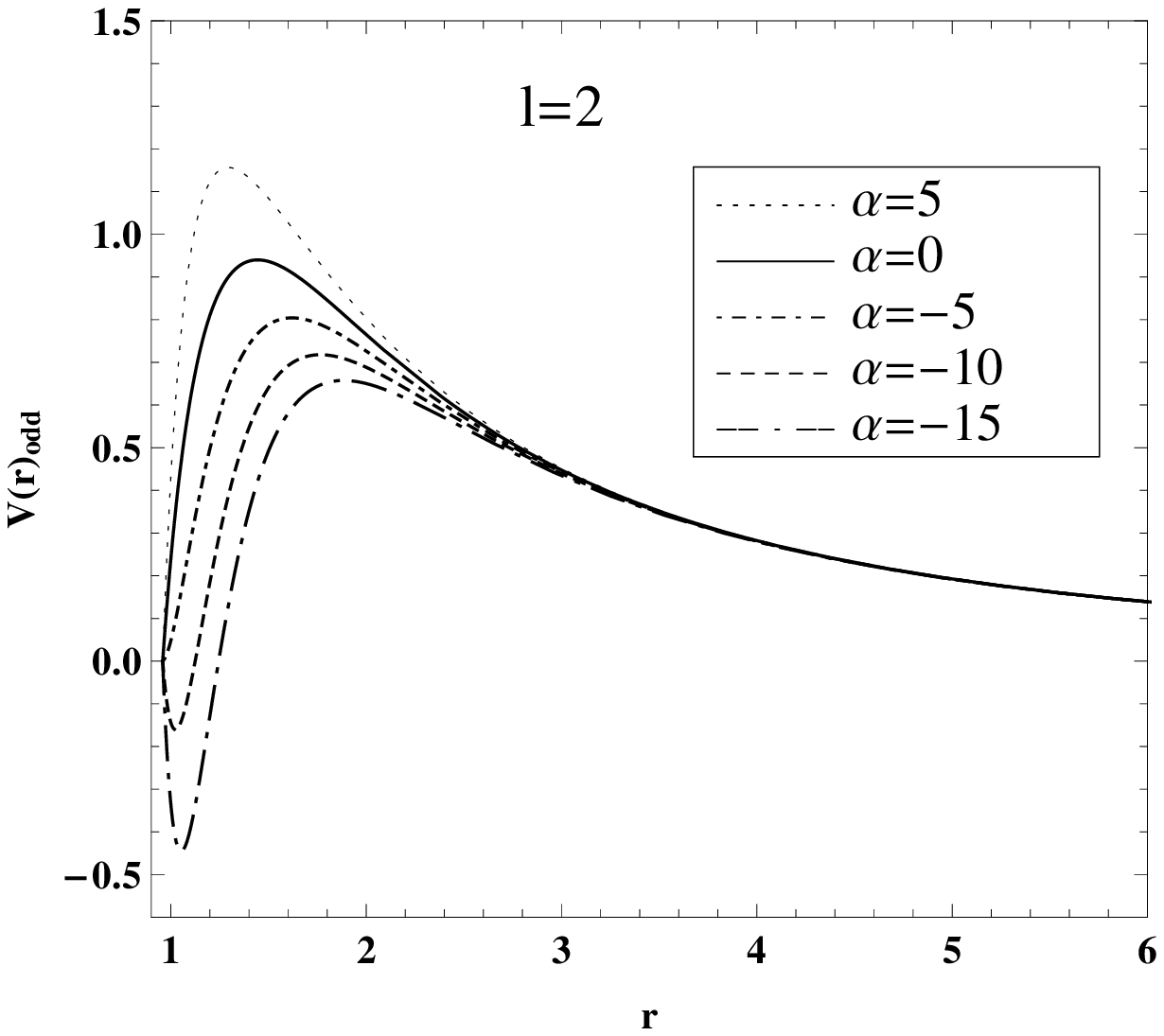}\includegraphics[width=5.6cm]{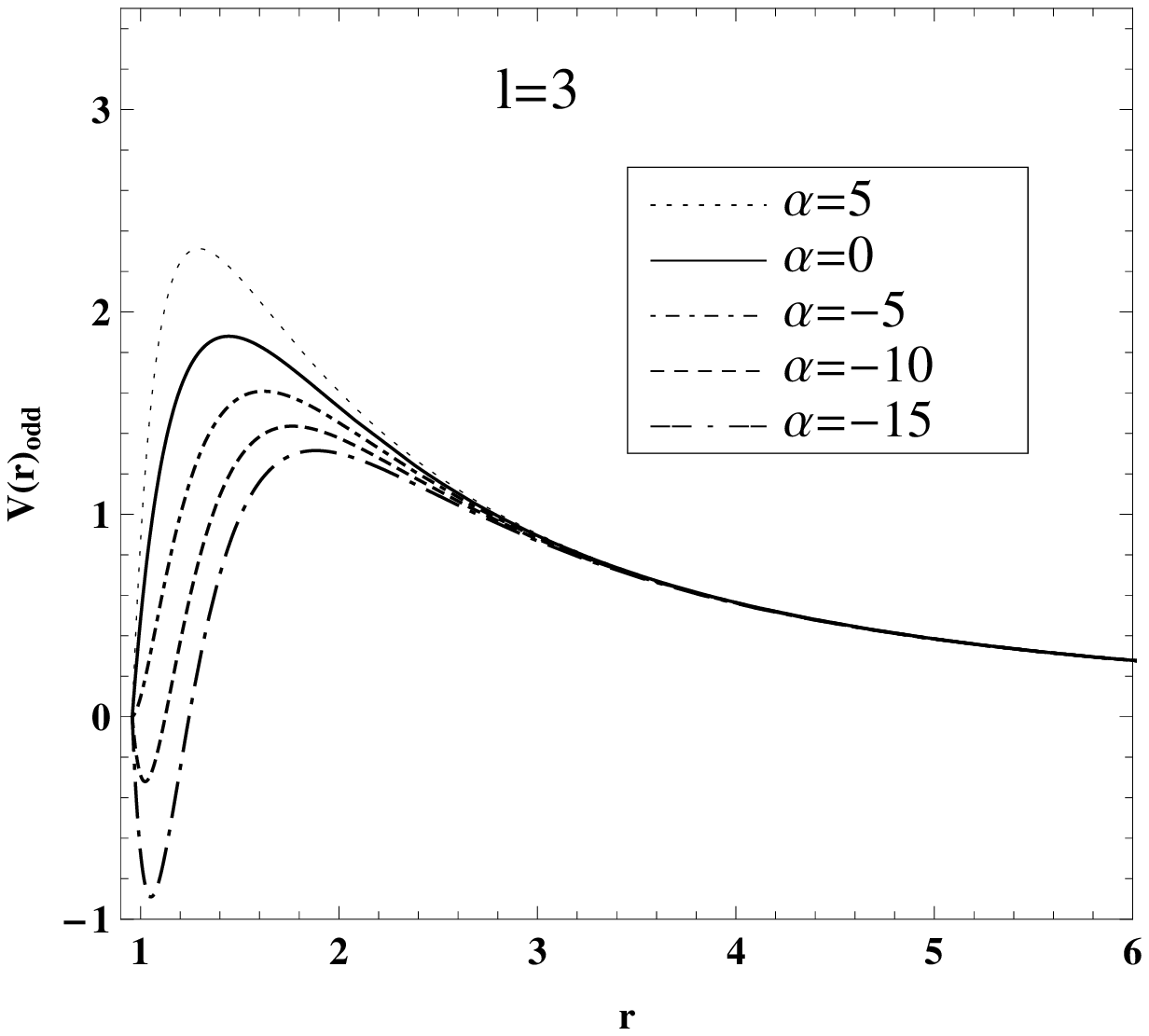}
\caption{Variation of the effective potential $V(r)_{odd}$ with the
polar coordinate $r$ for fixed $l=1$ (left), $l=2$ (middle) and
$l=3$ (right). The long-dash-dotted, dashed, short-dash-dotted, solid and
dotted  lines are corresponding to the cases with
$\alpha=-15,\;-10,\;-5,\;0,\;5$, respectively. We set $2M=1$ and $q=0.2$.}
\end{center}
\end{figure}
\begin{figure}[ht]
\begin{center}
\includegraphics[width=5.5cm]{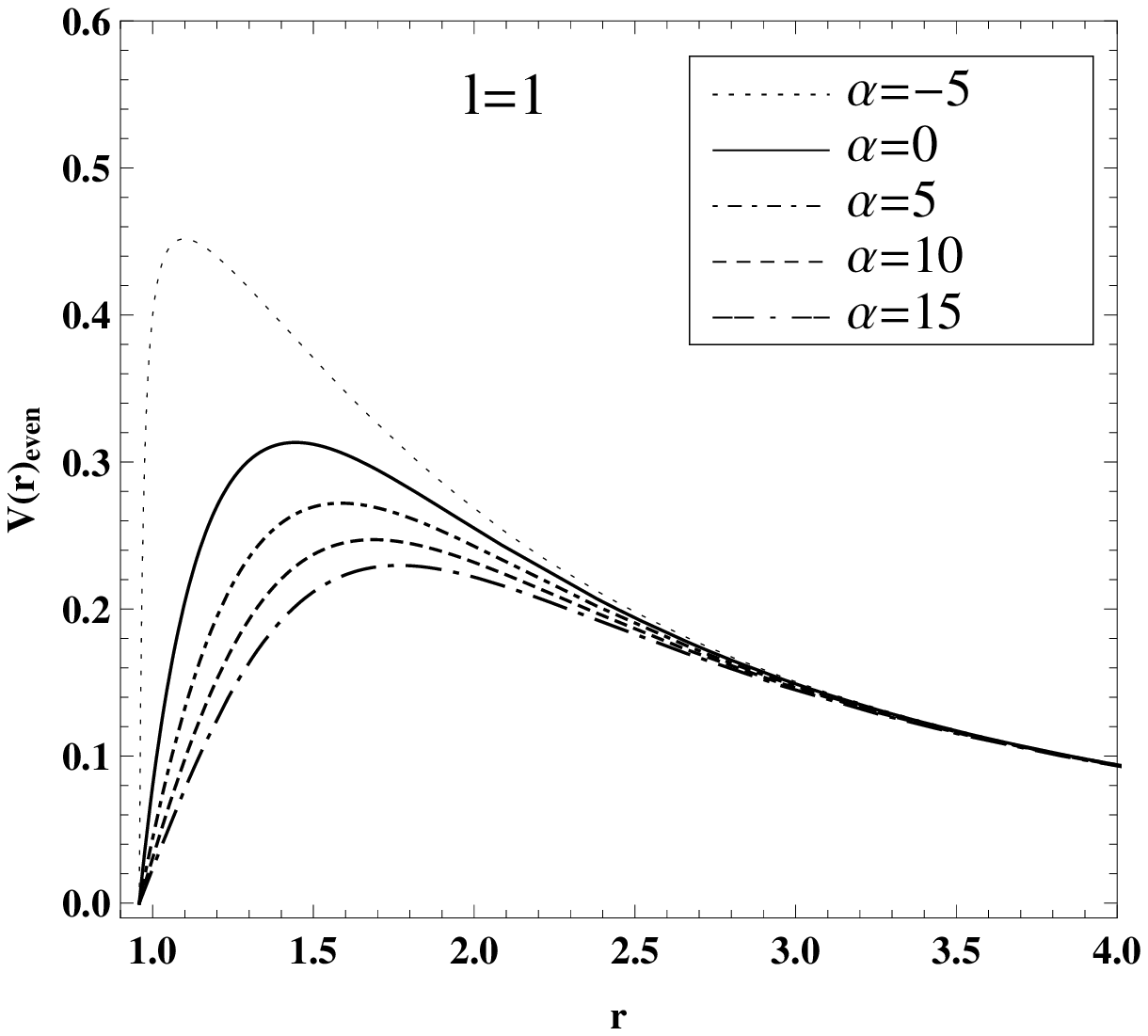}
\includegraphics[width=5.5cm]{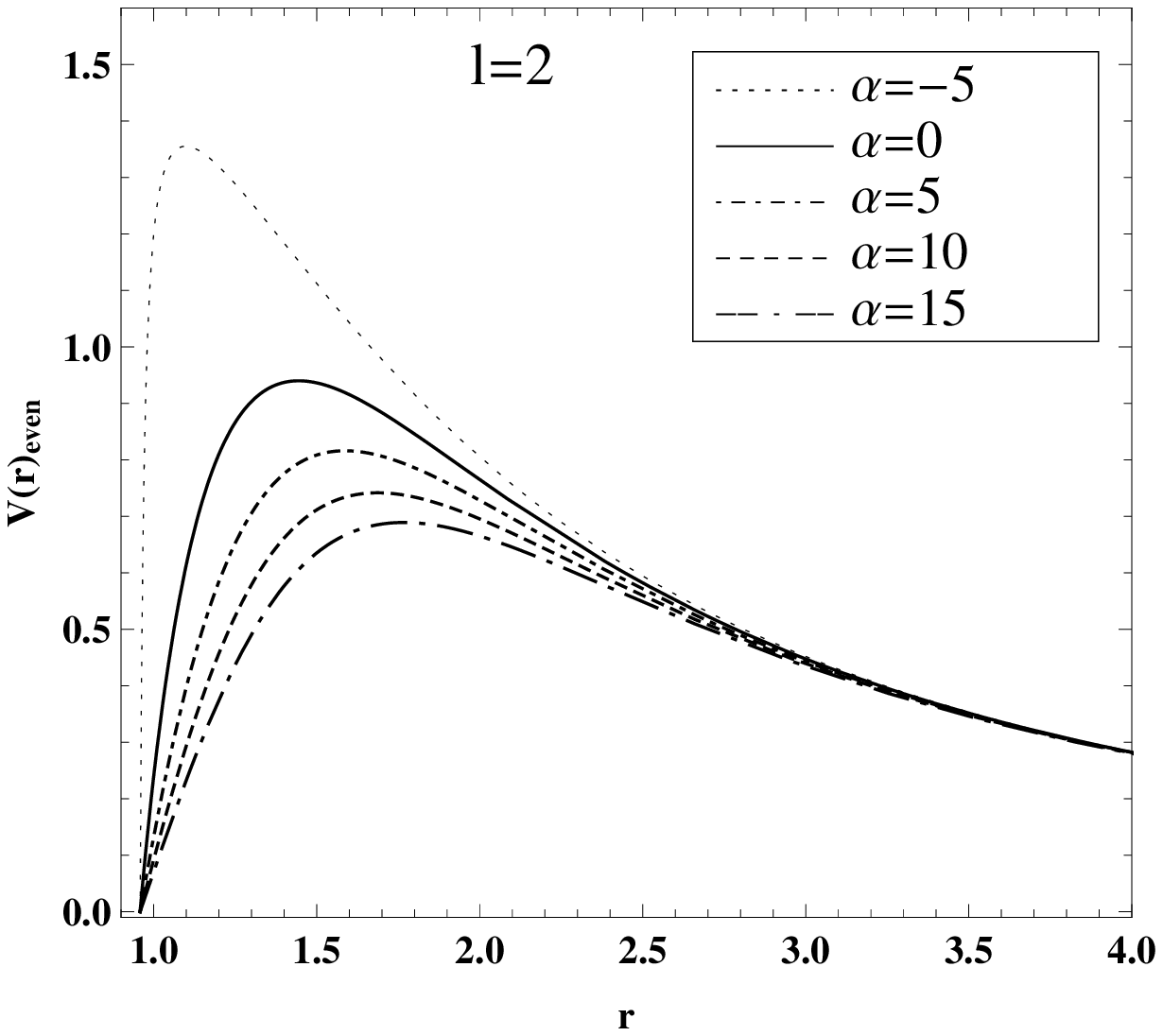}\includegraphics[width=5.6cm]{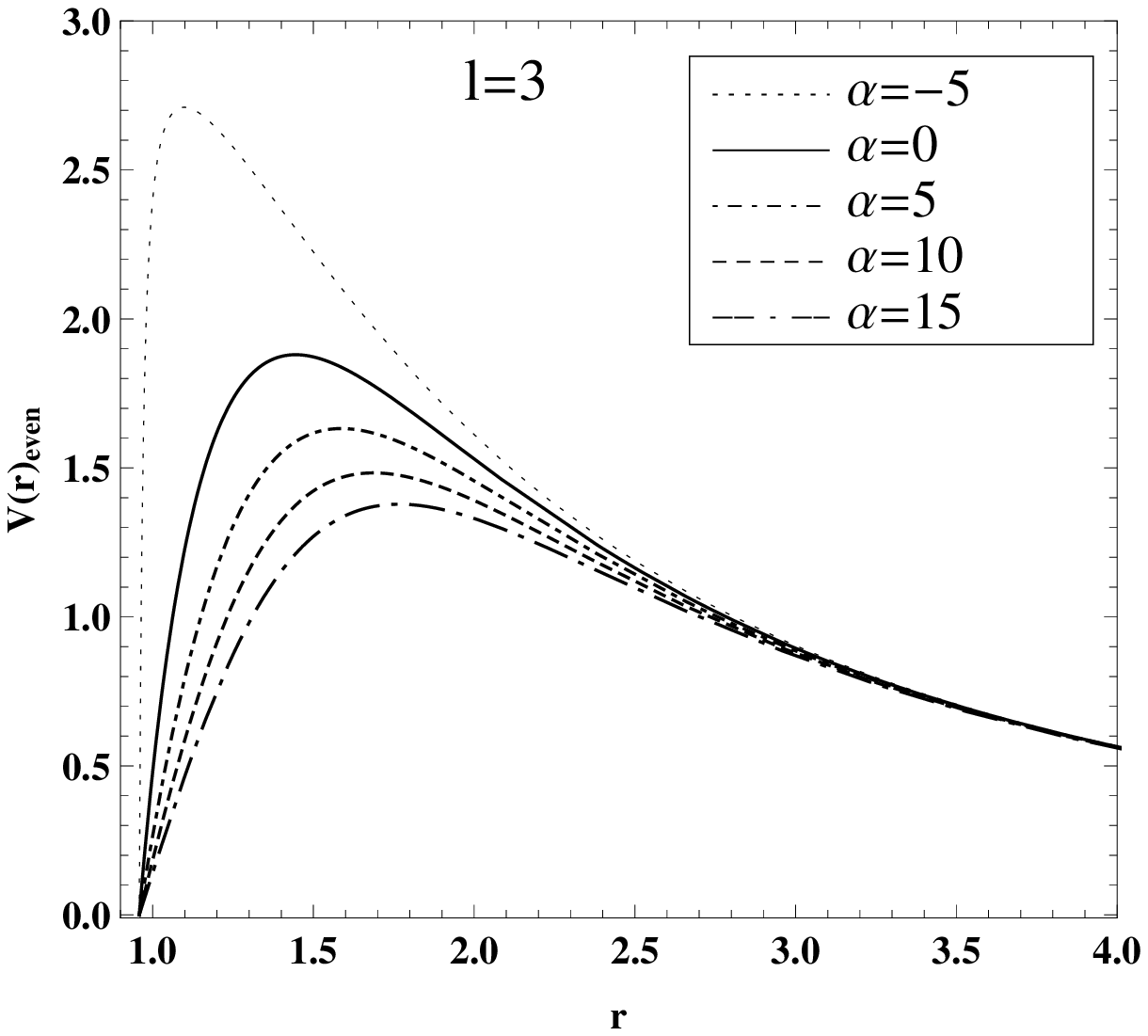}
\caption{Variation of the effective potential $V(r)_{even}$ with the
polar coordinate $r$ for fixed $l=1$ (left), $l=2$ (middle) and
$l=3$ (right). The long-dash-dotted, dashed, short-dash-dotted, solid and
dotted  lines are corresponding to the cases with
$\alpha=15,\;10,\;5,\;0,\;-5$, respectively. We set $2M=1$ and $q=0.2$.}
\end{center}
\end{figure}
The dependence of the effective potentials $V(r)_{odd}$ and $V(r)_{even}$ on
the coupling constant $\alpha$ for fixed $l$ is shown in Figs.(1) and (2).
For the effective potentials $V(r)_{odd}$, one can find that the peak height of the potential barrier increases with the
coupling constant $\alpha$ fixed $l$. For the effective potentials $V(r)_{even}$, we find that it decreases with the
coupling constant $\alpha$. These differences between the effective potentials $V(r)_{odd}$ and $V(r)_{even}$ imply that the features of the wave dynamics of the vector field perturbation coupling to Einstein tensor would depend heavily on the parities of the perturbation fields. Moreover, one
can find that in the allowed range of $\alpha$ both of $V(r)_{odd}$ with $\alpha\geq0$ and $V(r)_{even}$ are positive definite everywhere outside the black hole event
horizon. This implies that the solution of the wave equation (\ref{radial})
is bounded and the system is stable in this case. However, in the effective potentials $V(r)_{odd}$ with $\alpha<0$, there exits the negative
gap near the outer event horizon for the certain value of $\alpha$, and then the stability of the perturbational field is not guaranteed \cite{RAK1,RAK2,Card}. In the following section, we will check the behavior of the external vector field perturbation coupling to Einstein tensor in the background of a Reissner-Nordstr\"{o}m black hole if the coupling constant $\alpha$ lies in the region where the negative gap appears in the effective potentials.

In order to investigate the dependence of the quasinormal modes  on the coupling constant $\alpha$, we calculate the fundamental quasinormal modes ($n=0$) of two vector field perturbations with different parities by using the WKB approximation method \cite{Schutz:1985,Iyer:1987}. For the quasinormal modes,  we have the wave function $\Psi(r,t)=\Psi(r)e^{-i\omega t}$ with $\omega =\omega_{R}+i\omega_{I}$.  The real part $\omega_{R}$ and the imaginary part $\omega_{I}$  denote the frequency and the change rate of amplitude of the oscillation, respectively. The modes with $\omega_{I}<0$ corresponds to a stable  mode and the negative value of $\omega_{I}$ reflects the decay rate of the mode. The fundamental quasinormal modes decays more slowly than other overtone modes ($n>0$).
The change of the fundamental quasinormal frequencies of vector field perturbation with the coupling constant $\alpha$ and the charge $q$ is plotted in Figs. (3)-(6).
We find that the real parts of fundamental quasinormal frequencies of the odd parity perturbation increase monotonously with the coupling constant $\alpha$ for arbitrary $q$. This means that the vibration of the odd parity vector perturbation becomes faster for increasing values of $\alpha$.
For the even parity perturbation, one can obtain that the change of the real parts with $\alpha$ is opposite to that for the odd parity perturbation, which means that the vibration of the even parity vector perturbation becomes more slowly with the increasing $\alpha$.
\begin{figure}
\begin{center}
\includegraphics[width=5.5cm]{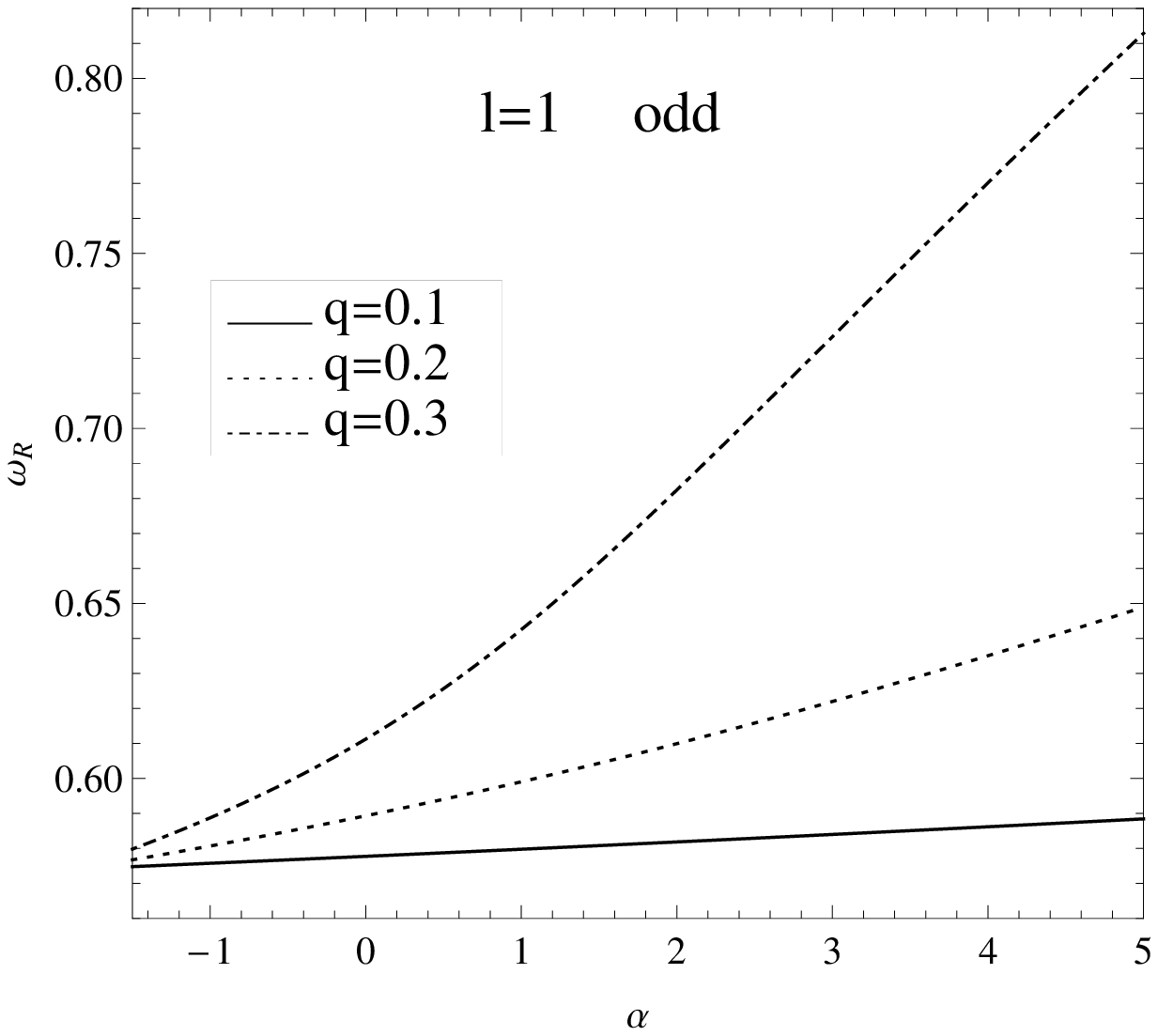}
\includegraphics[width=5.5cm]{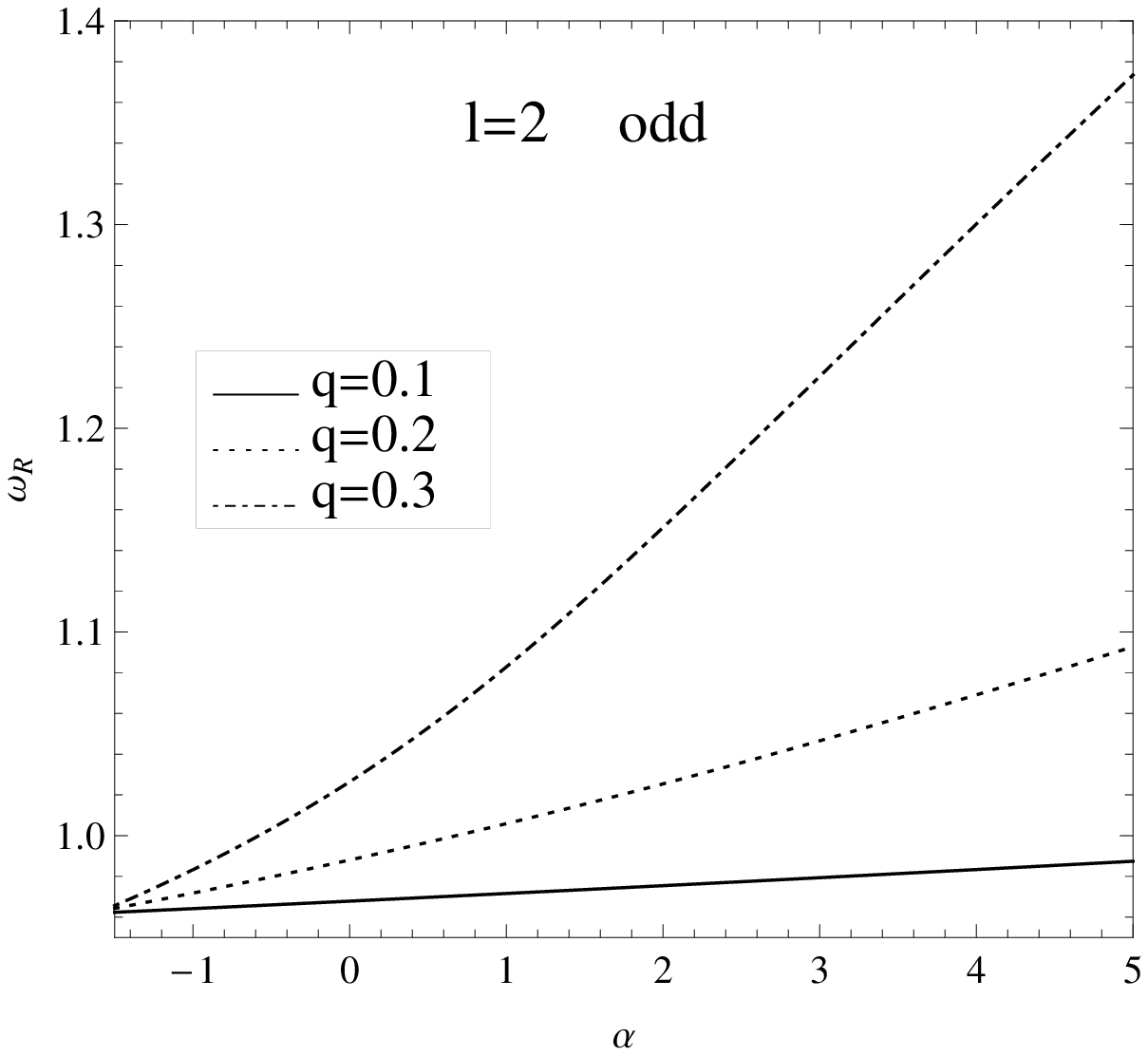}
\includegraphics[width=5.5cm]{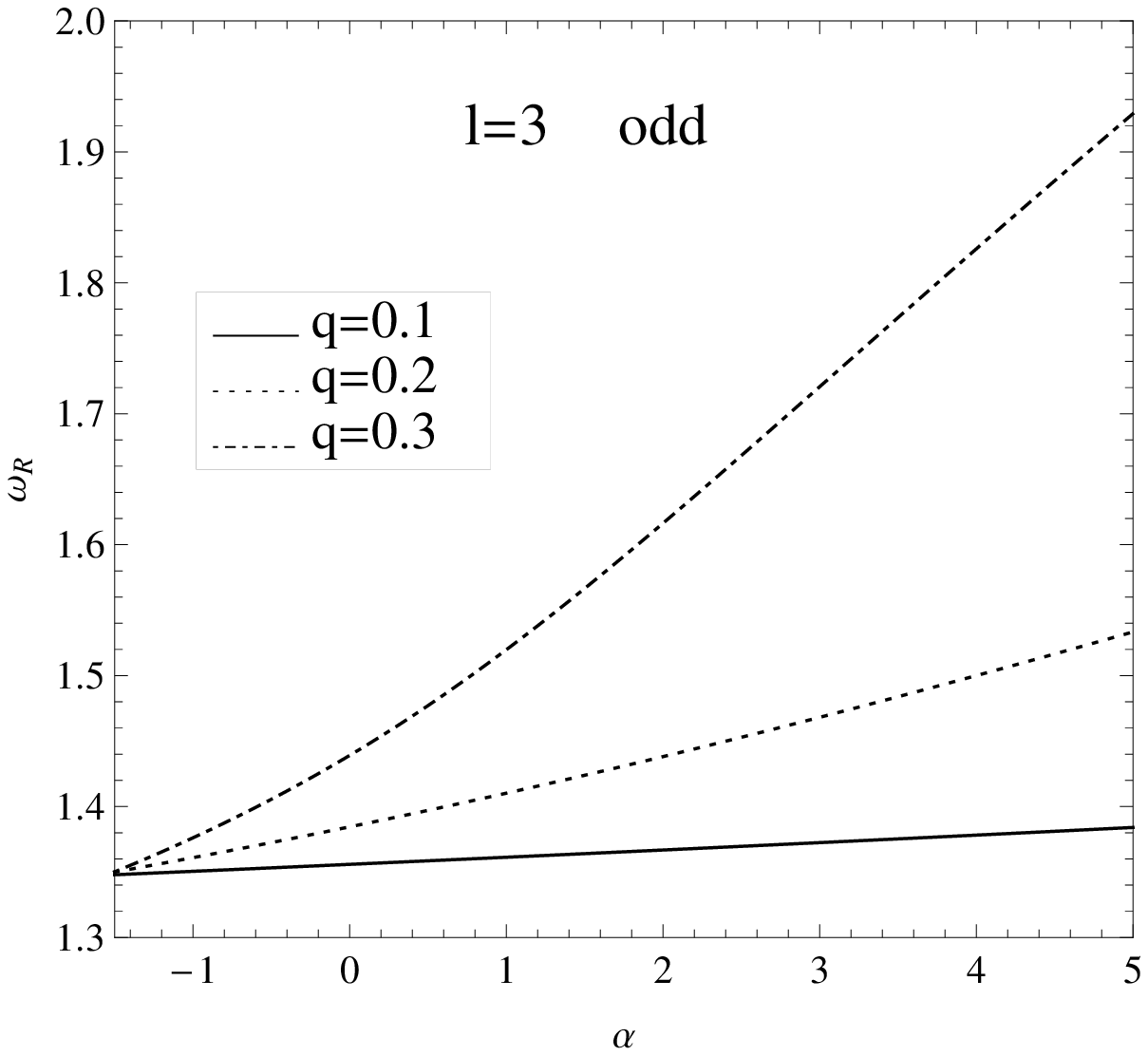}\\
\includegraphics[width=5.5cm]{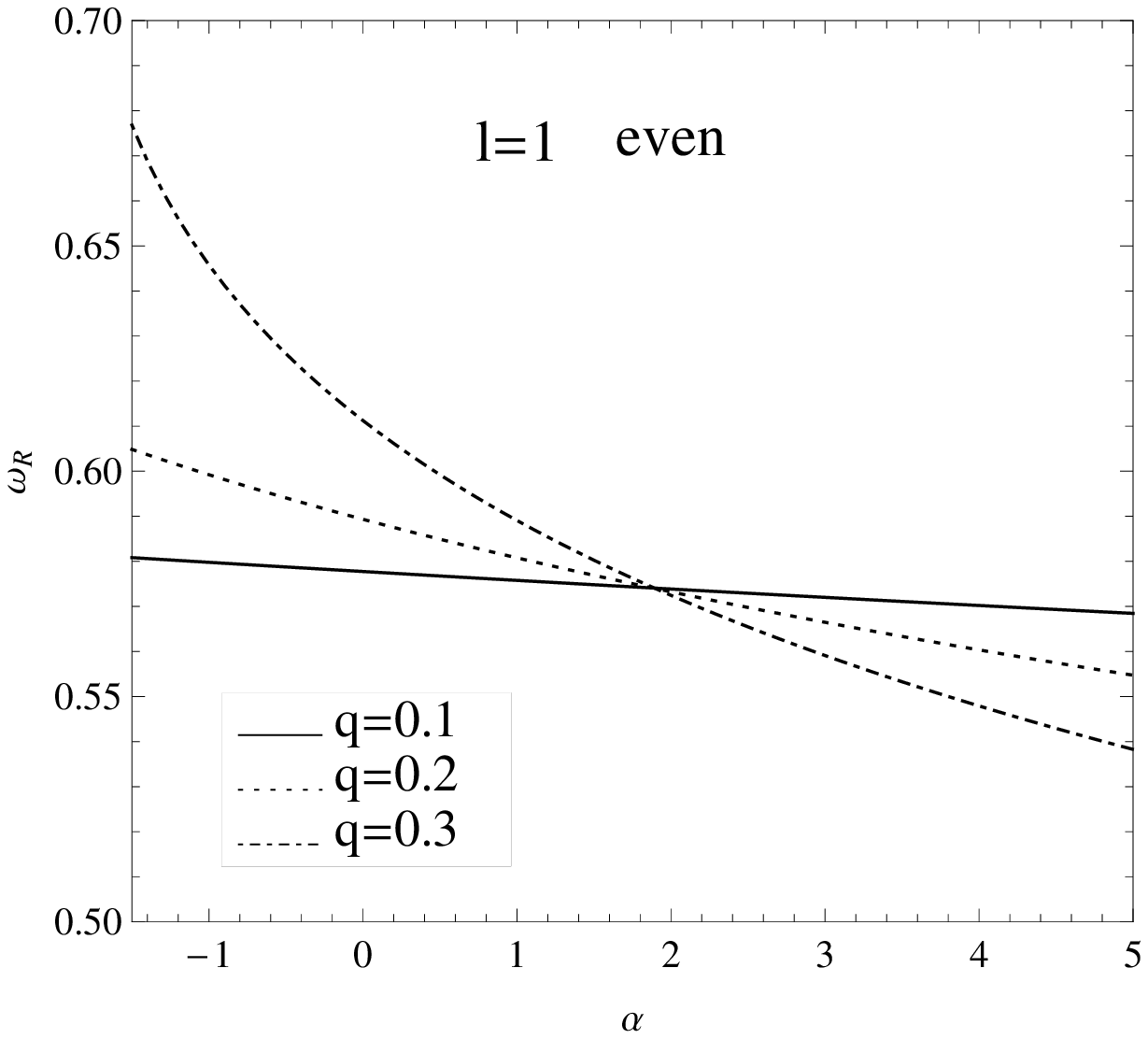}
\includegraphics[width=5.5cm]{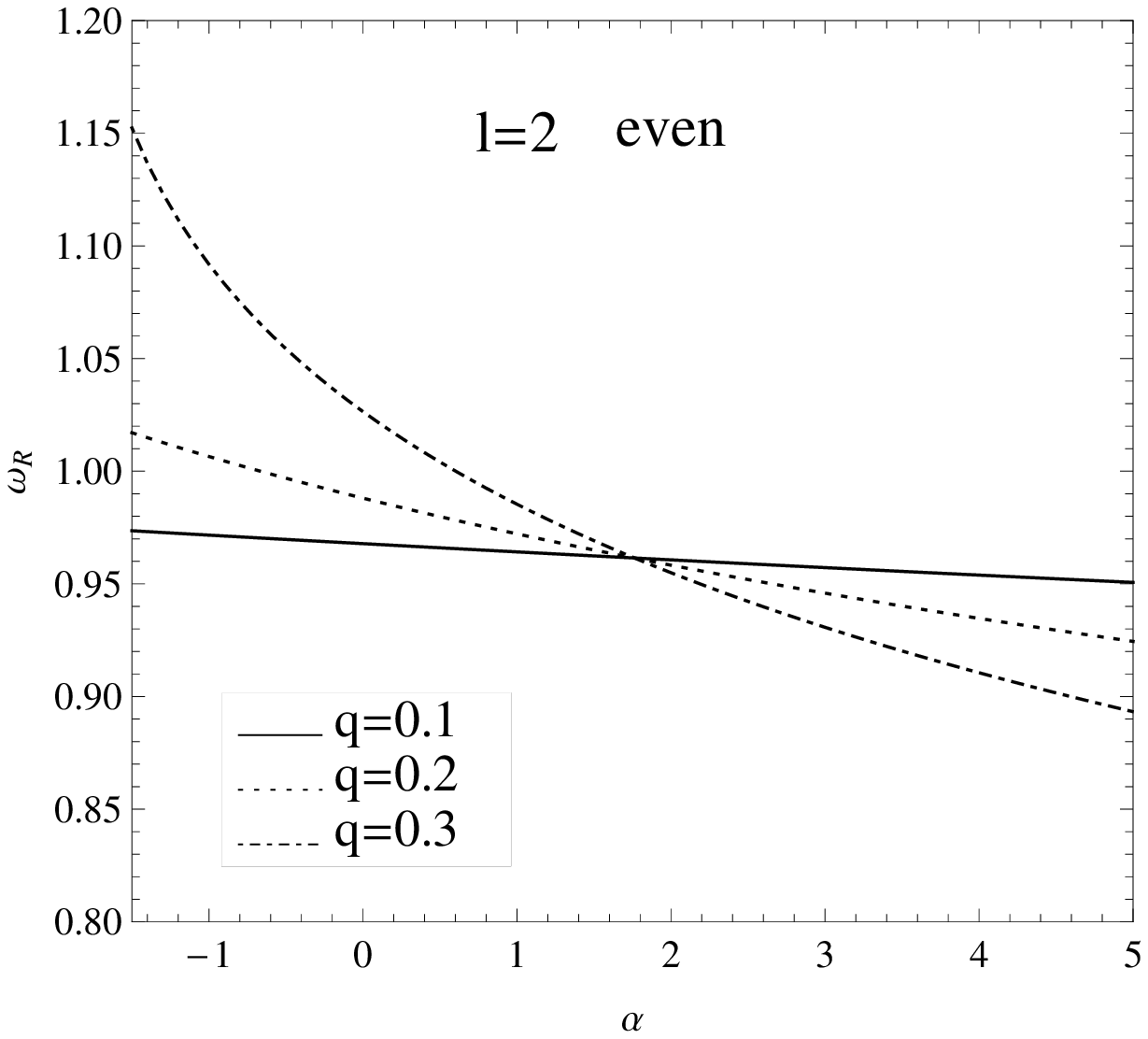}
\includegraphics[width=5.5cm]{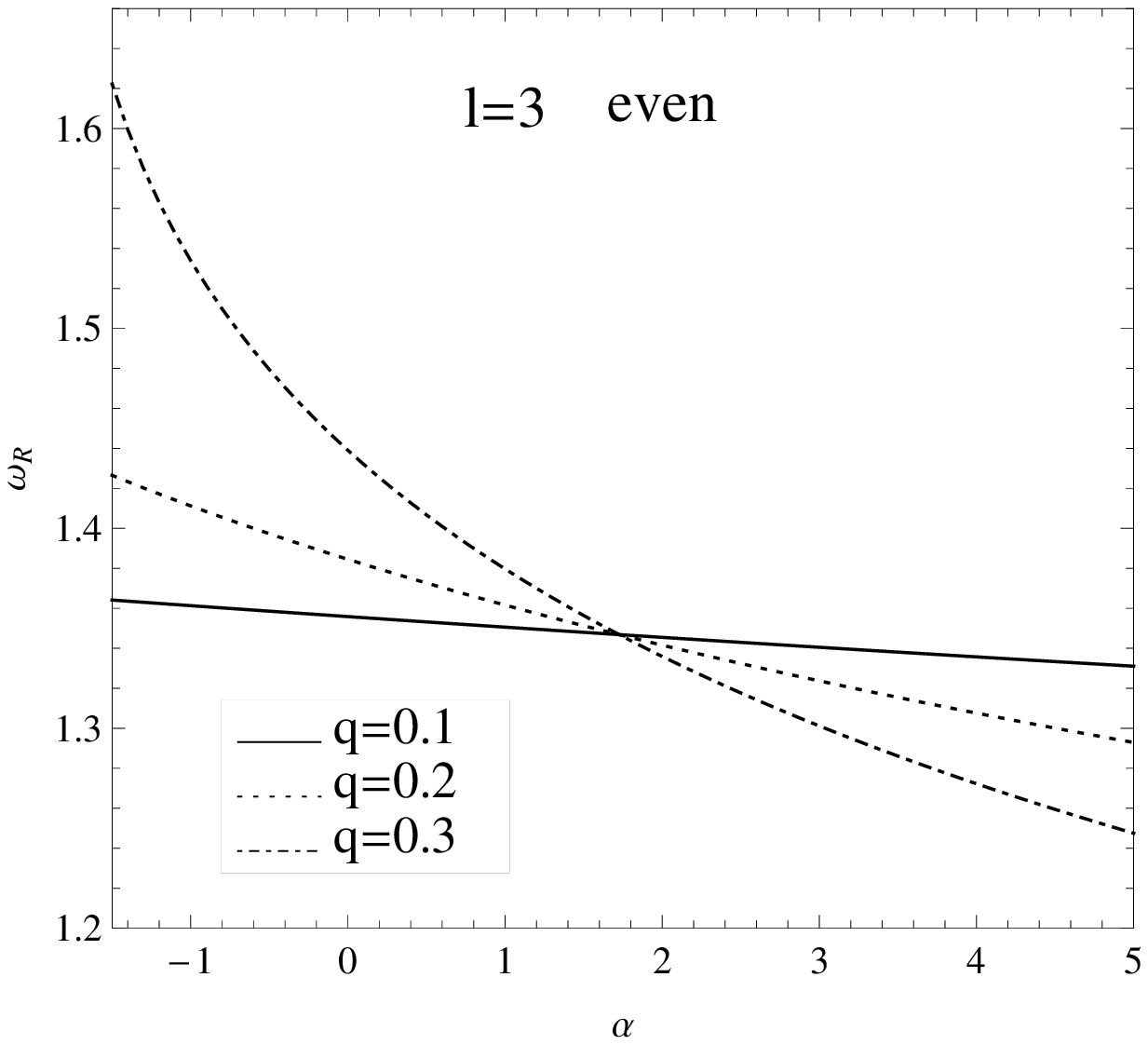}
\caption{Effects of the coupling parameter $\alpha$ on the real parts of the fundamental quasinormal modes of vector field perturbation with
the odd parity(the top row) or the even parity (the bottom row)  in
the Reissner-Nordstr\"{o}m black hole spacetime for fixed $q$. We set $2M=1$.}
\end{center}
\end{figure}
\begin{figure}
\begin{center}
\includegraphics[width=5.5cm]{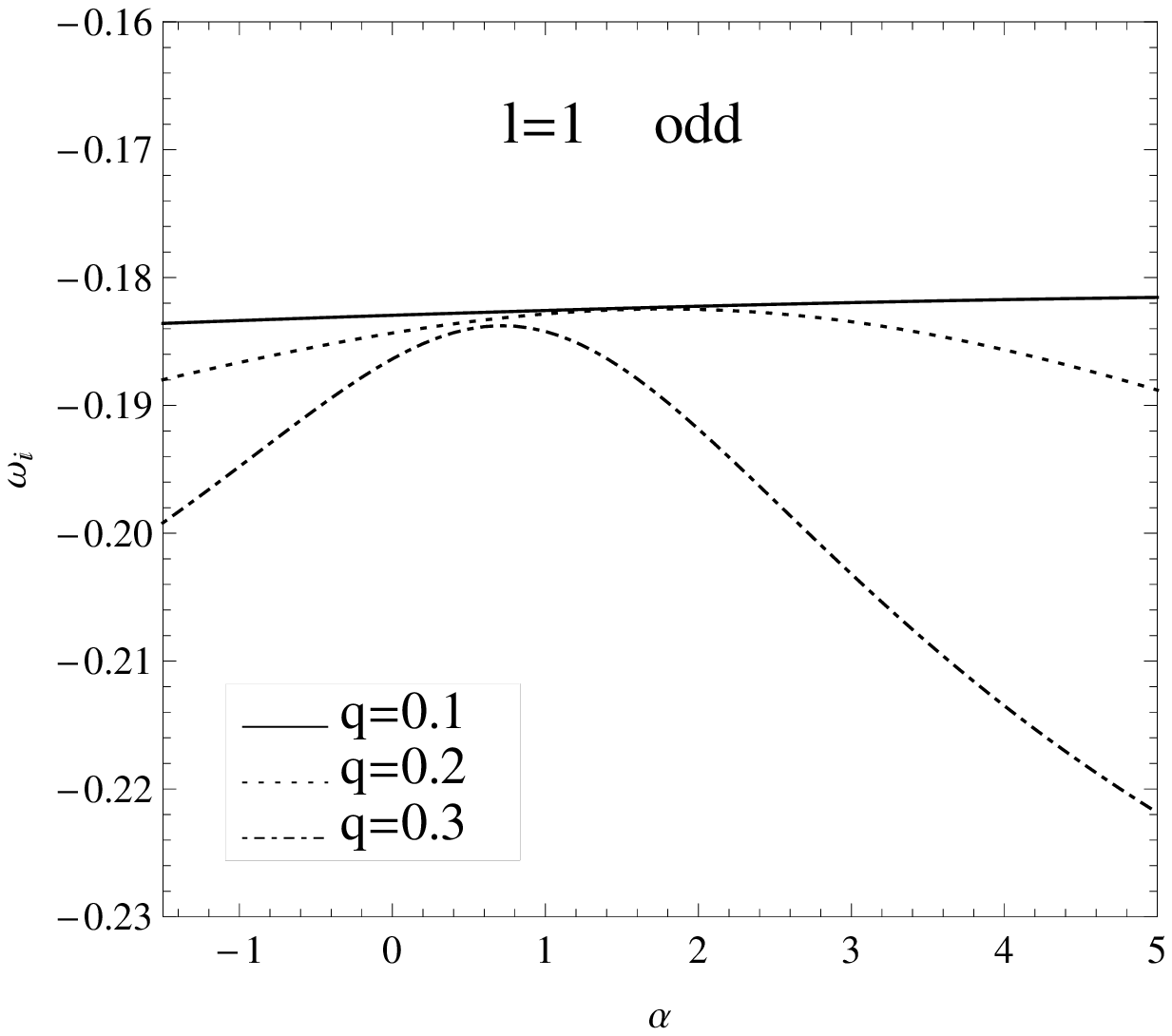}
\includegraphics[width=5.5cm]{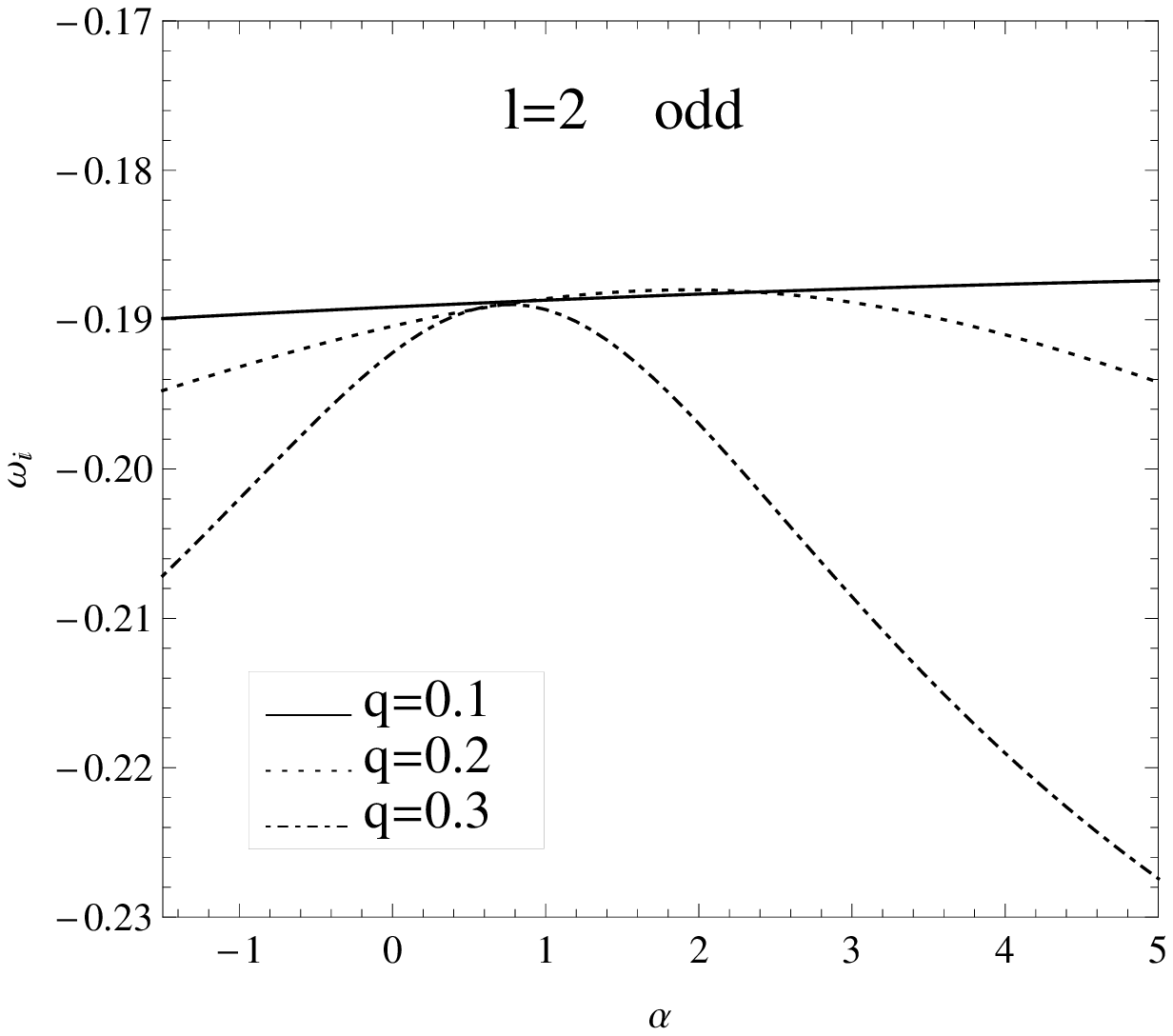}
\includegraphics[width=5.5cm]{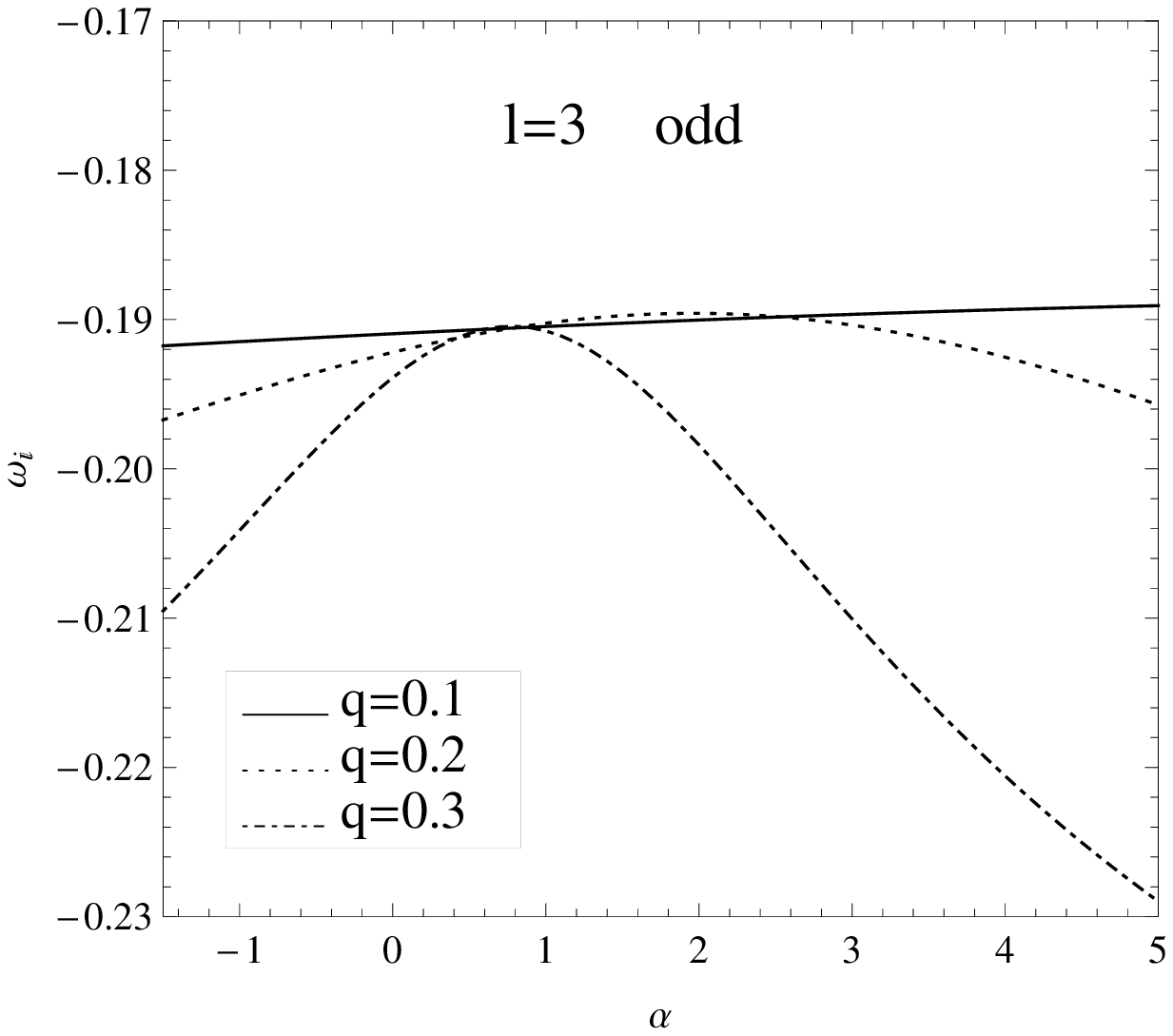}\\
\includegraphics[width=5.5cm]{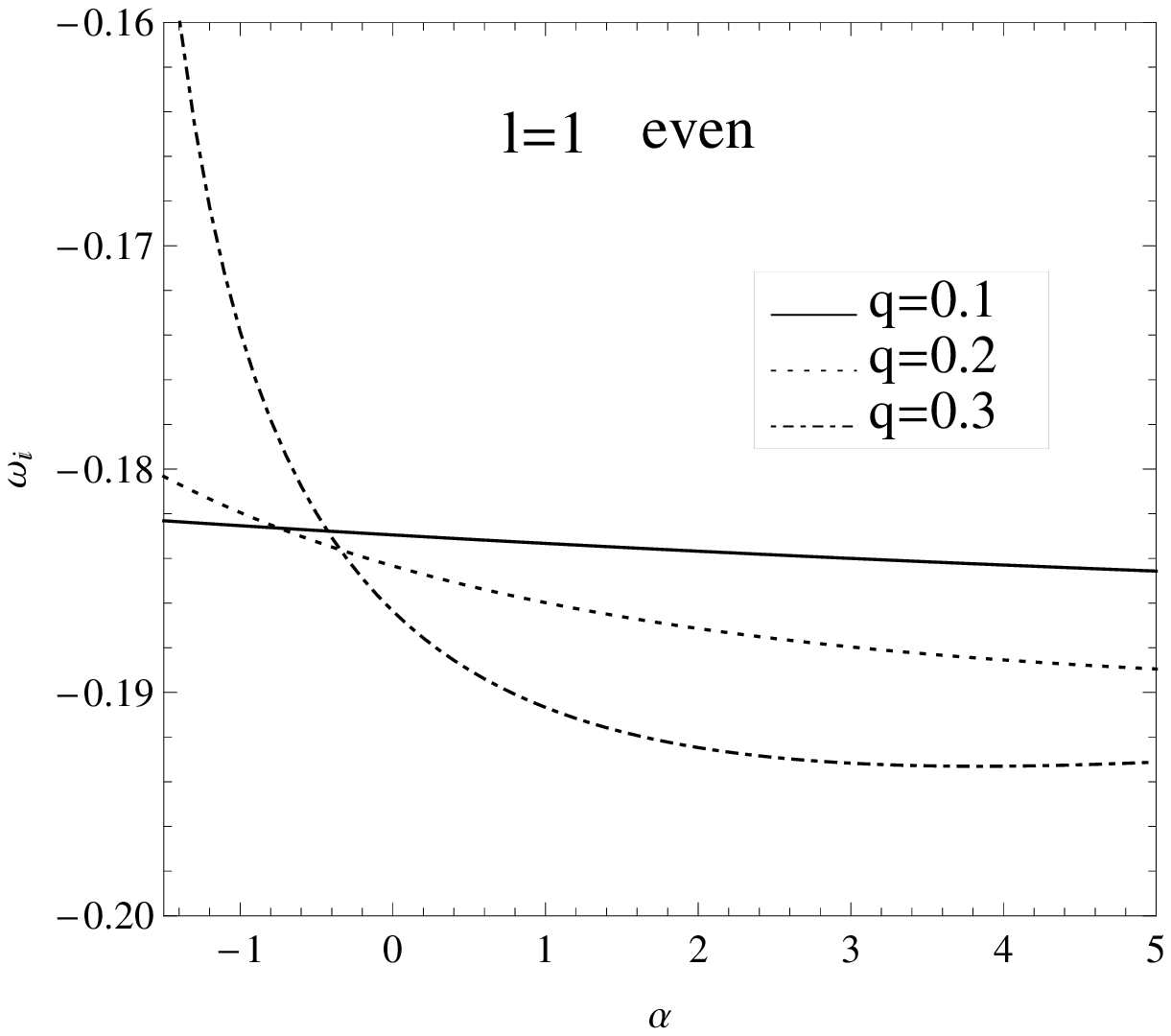}
\includegraphics[width=5.5cm]{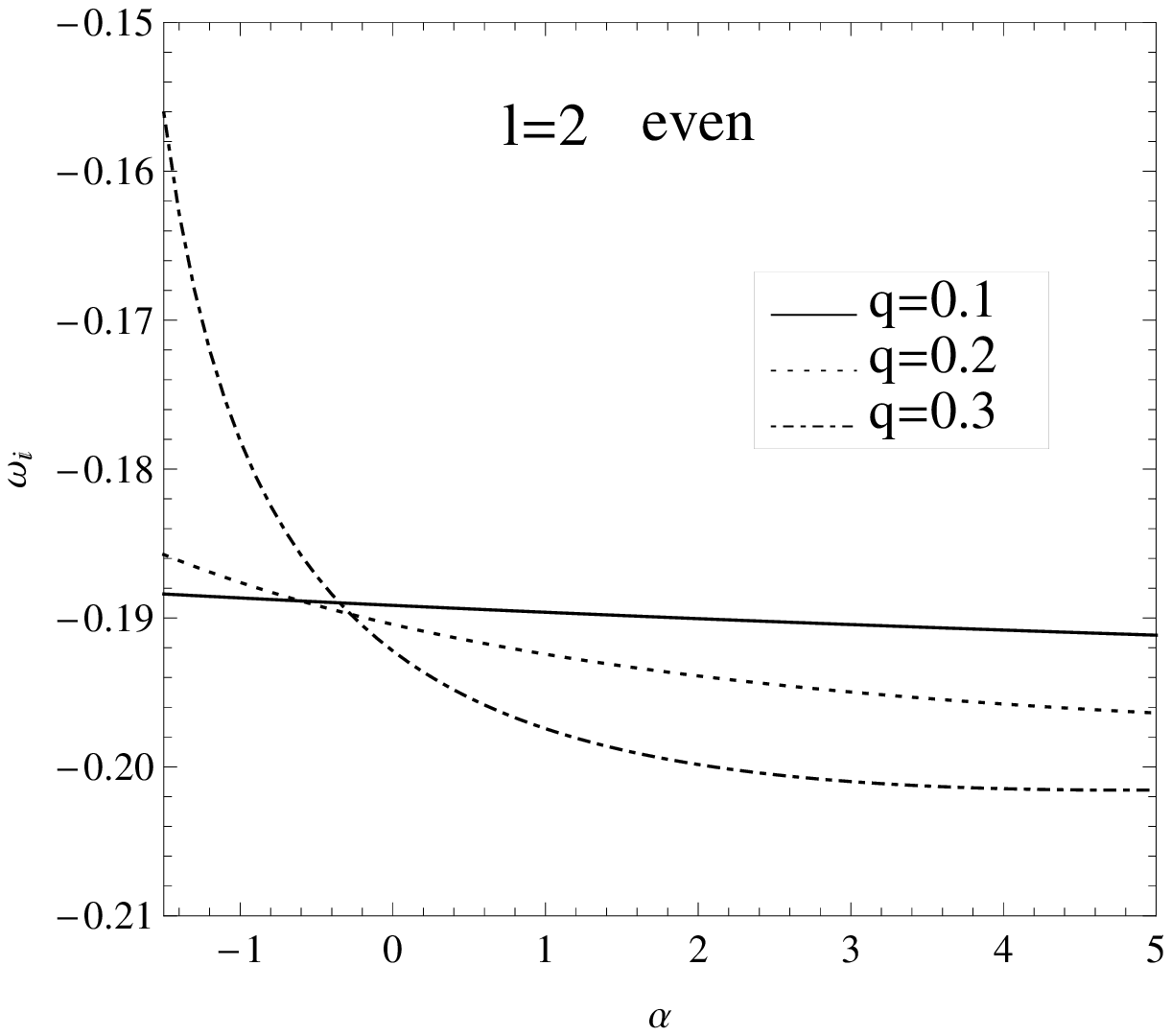}
\includegraphics[width=5.5cm]{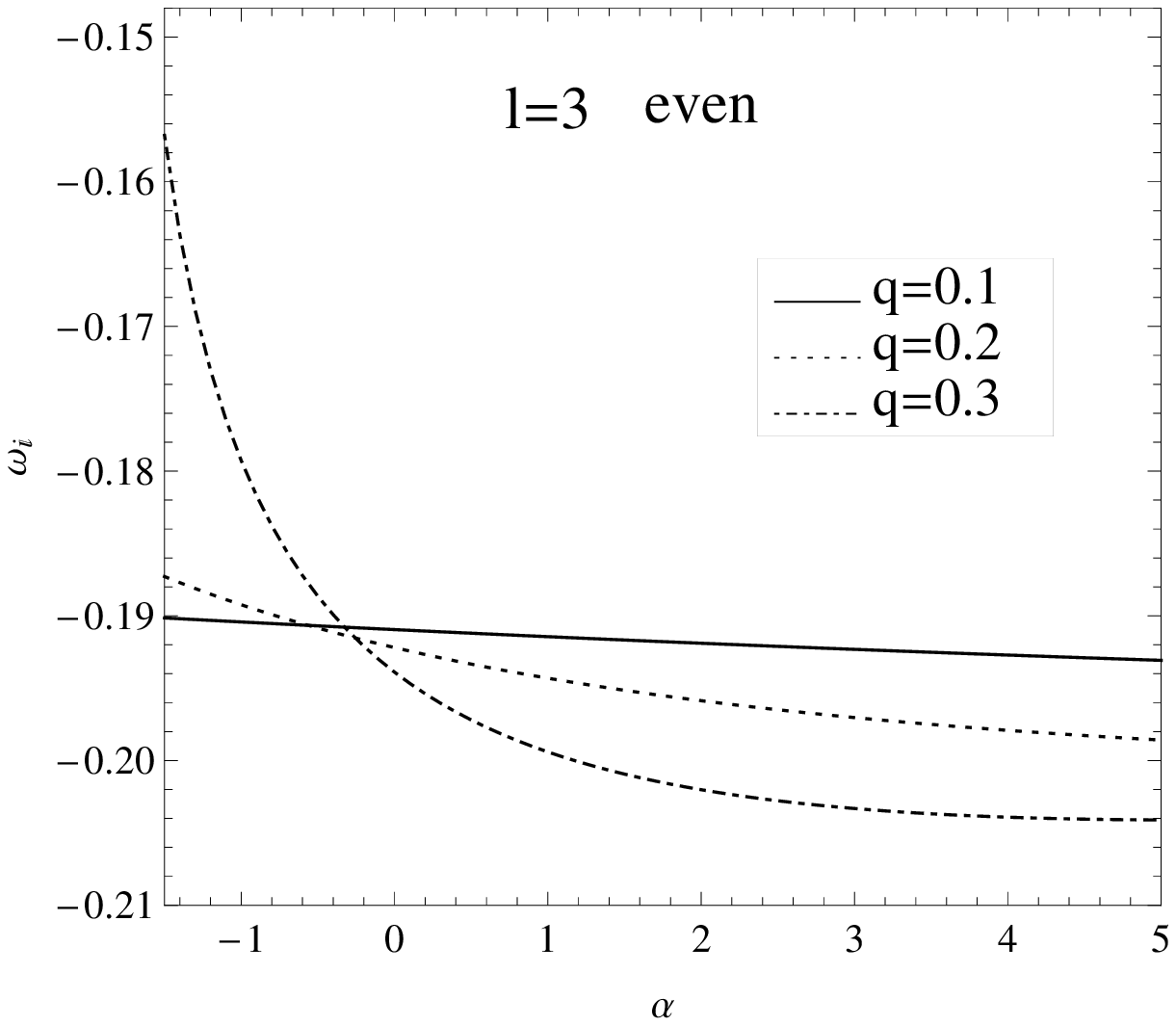}
\caption{Effects of the coupling parameter $\alpha$ on the imaginary parts of the fundamental quasinormal modes of vector field perturbation with
the odd parity(the top row) or the even parity (the bottom row)  in
the Reissner-Nordstr\"{o}m black hole spacetime for fixed $q$. We set $2M=1$.}
\end{center}
\end{figure}
\begin{figure}
\begin{center}
\includegraphics[width=5.5cm]{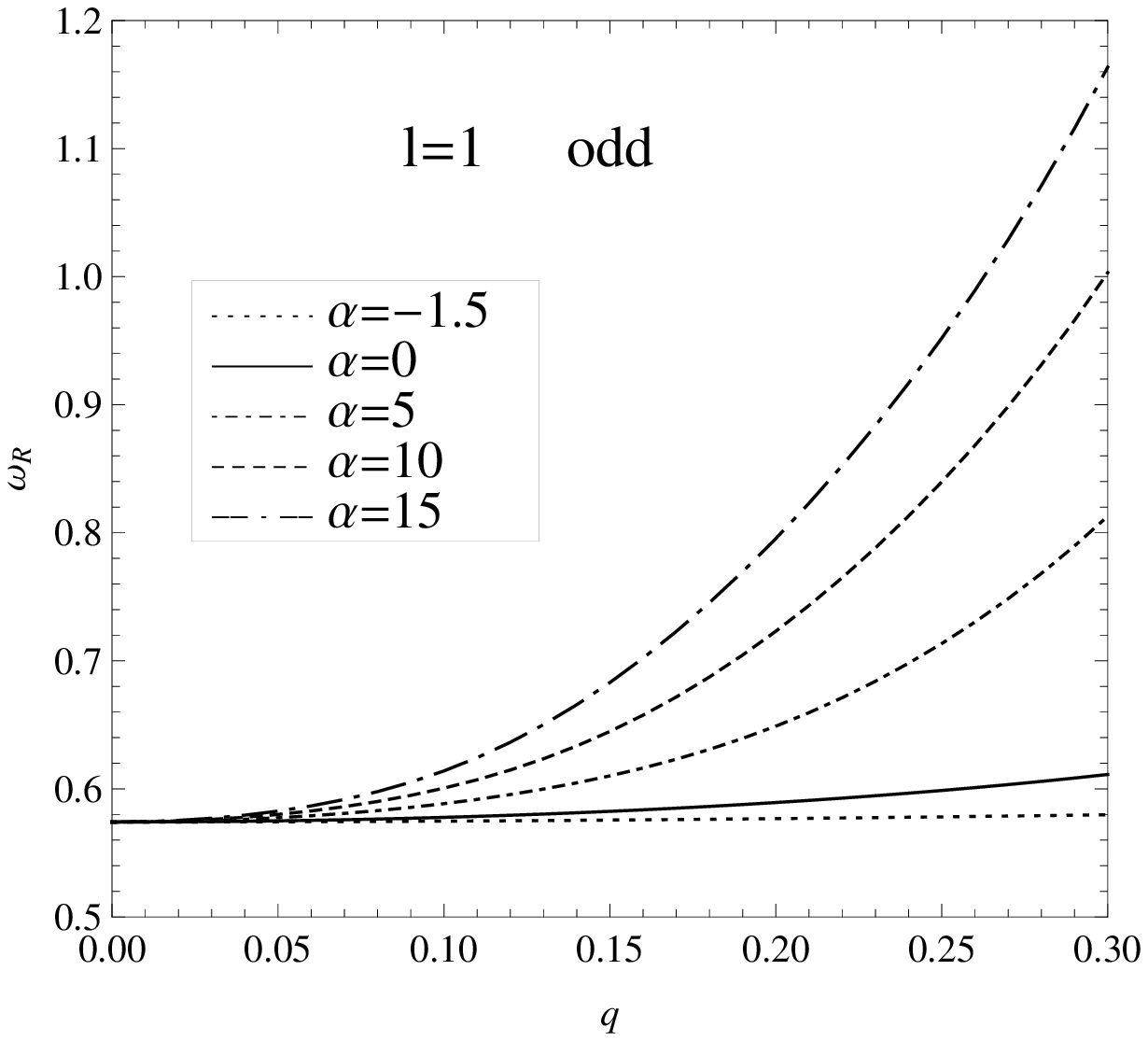}
\includegraphics[width=5.5cm]{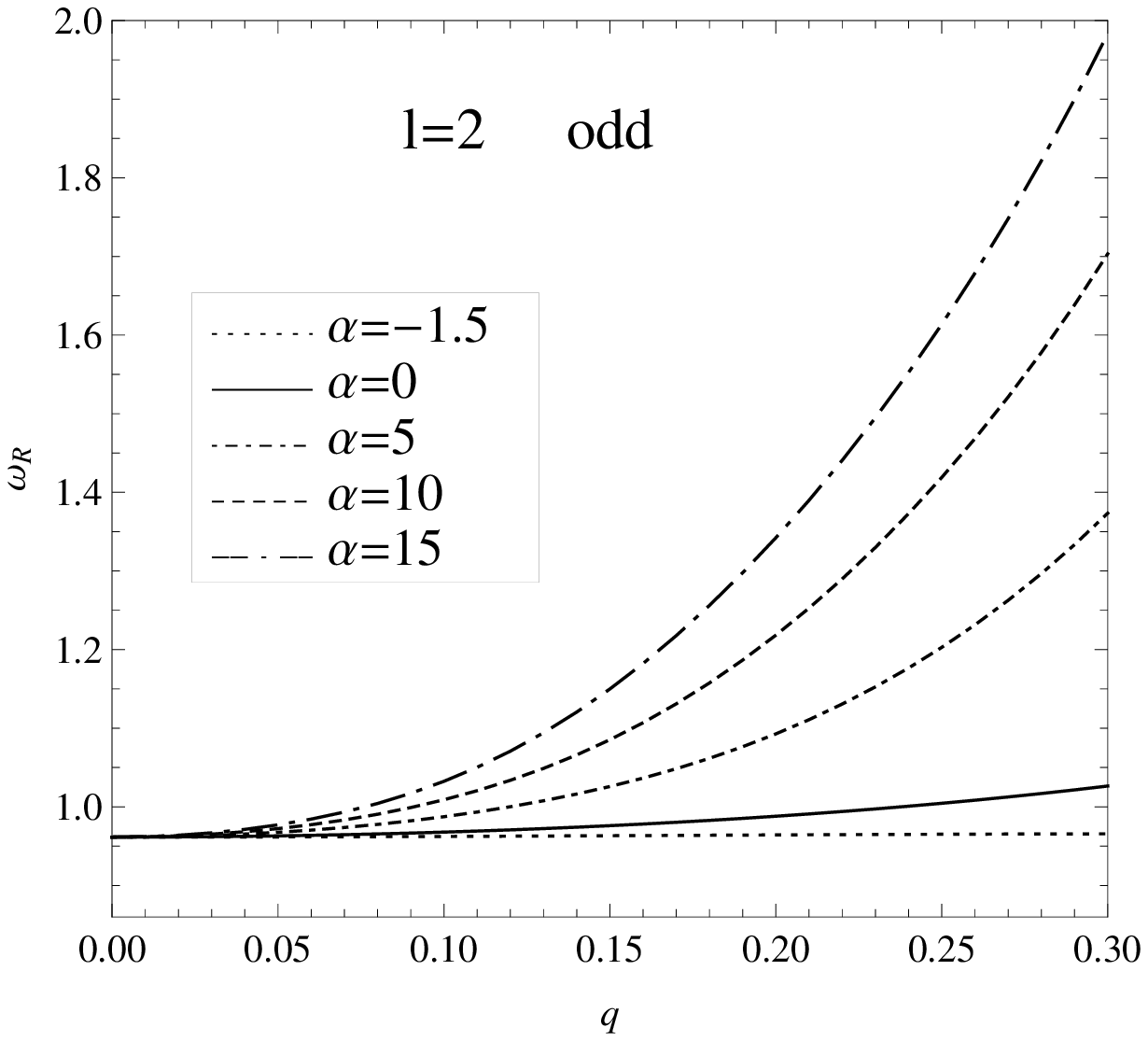}
\includegraphics[width=5.5cm]{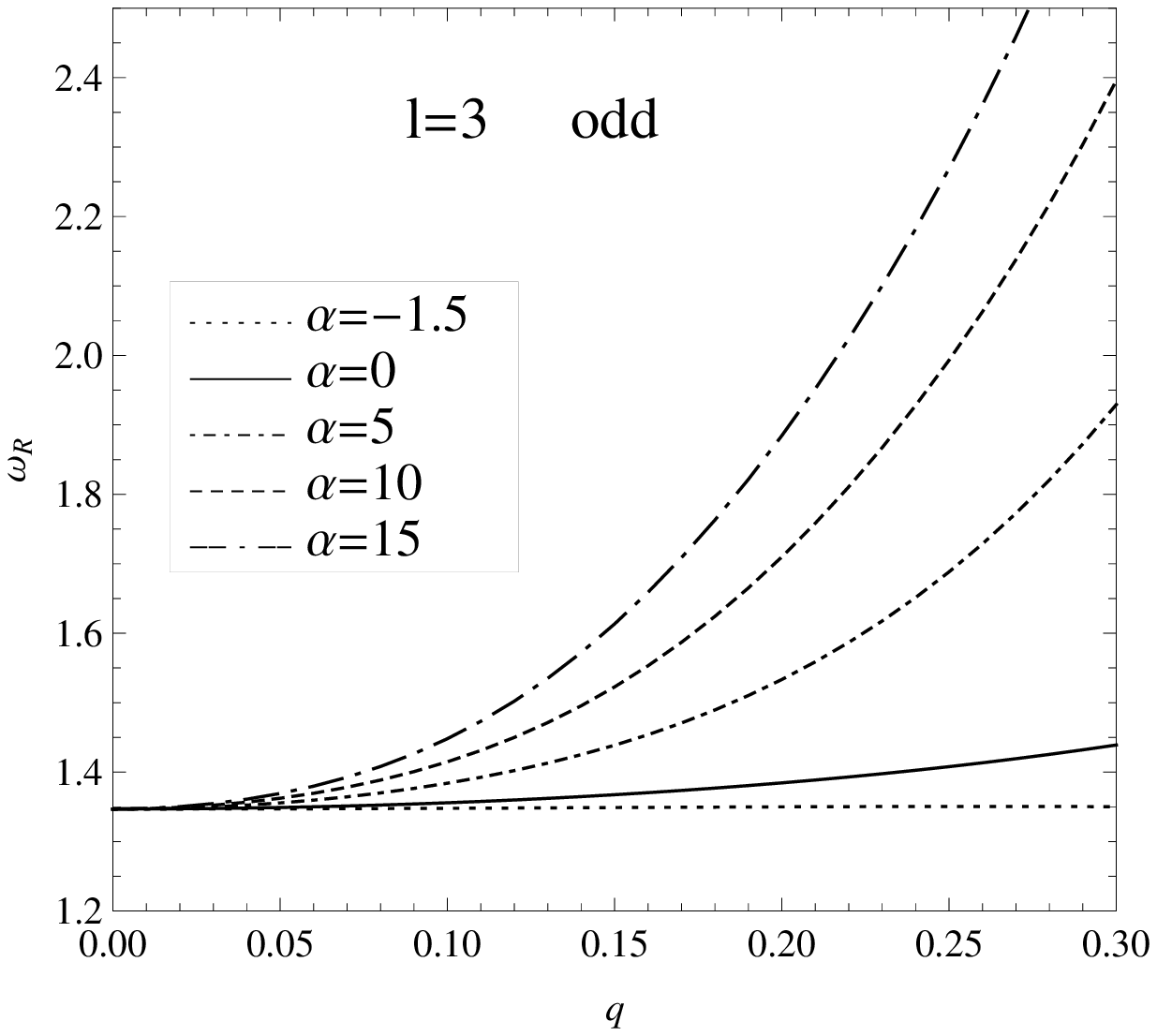}\\
\includegraphics[width=5.5cm]{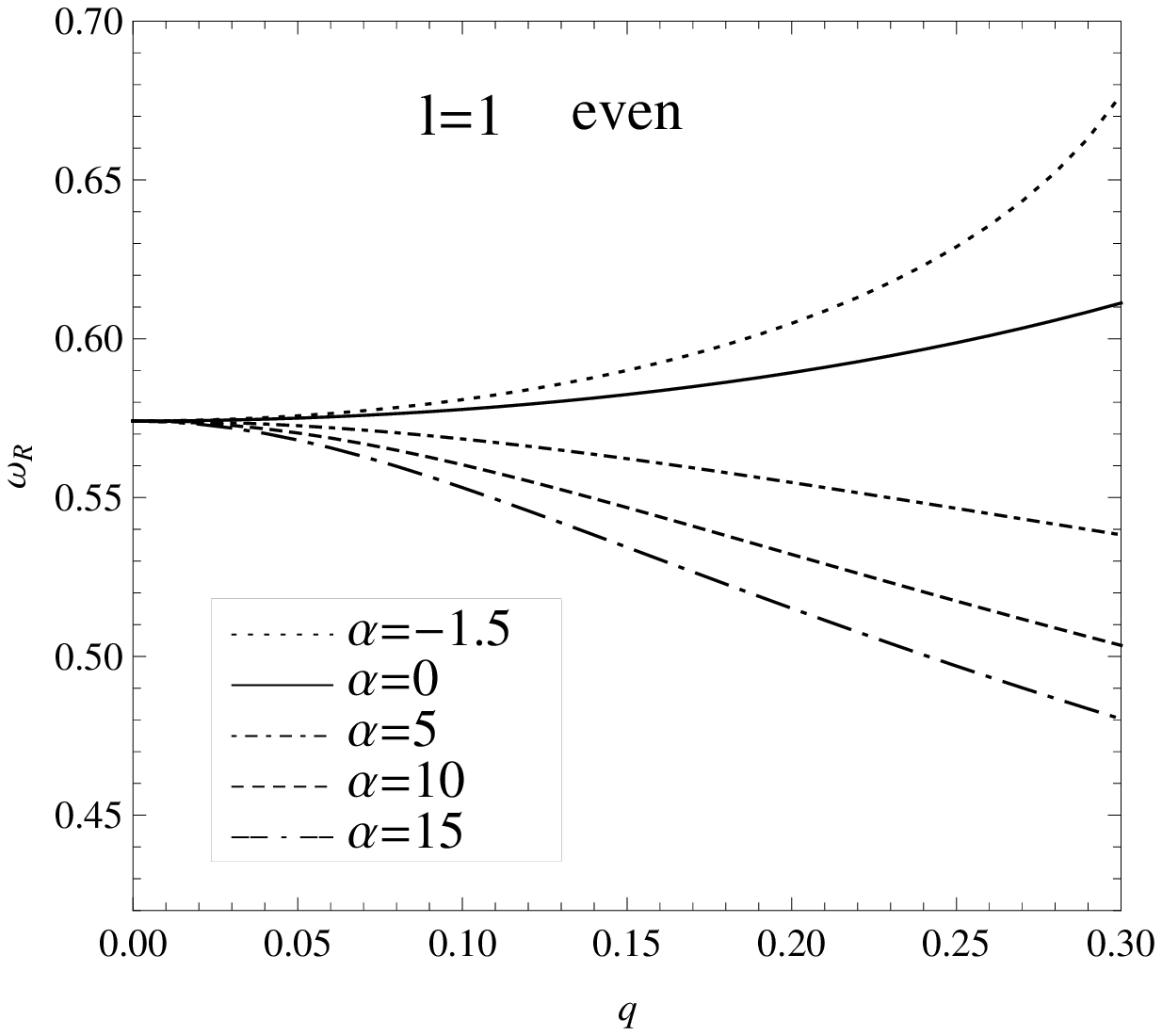}
\includegraphics[width=5.5cm]{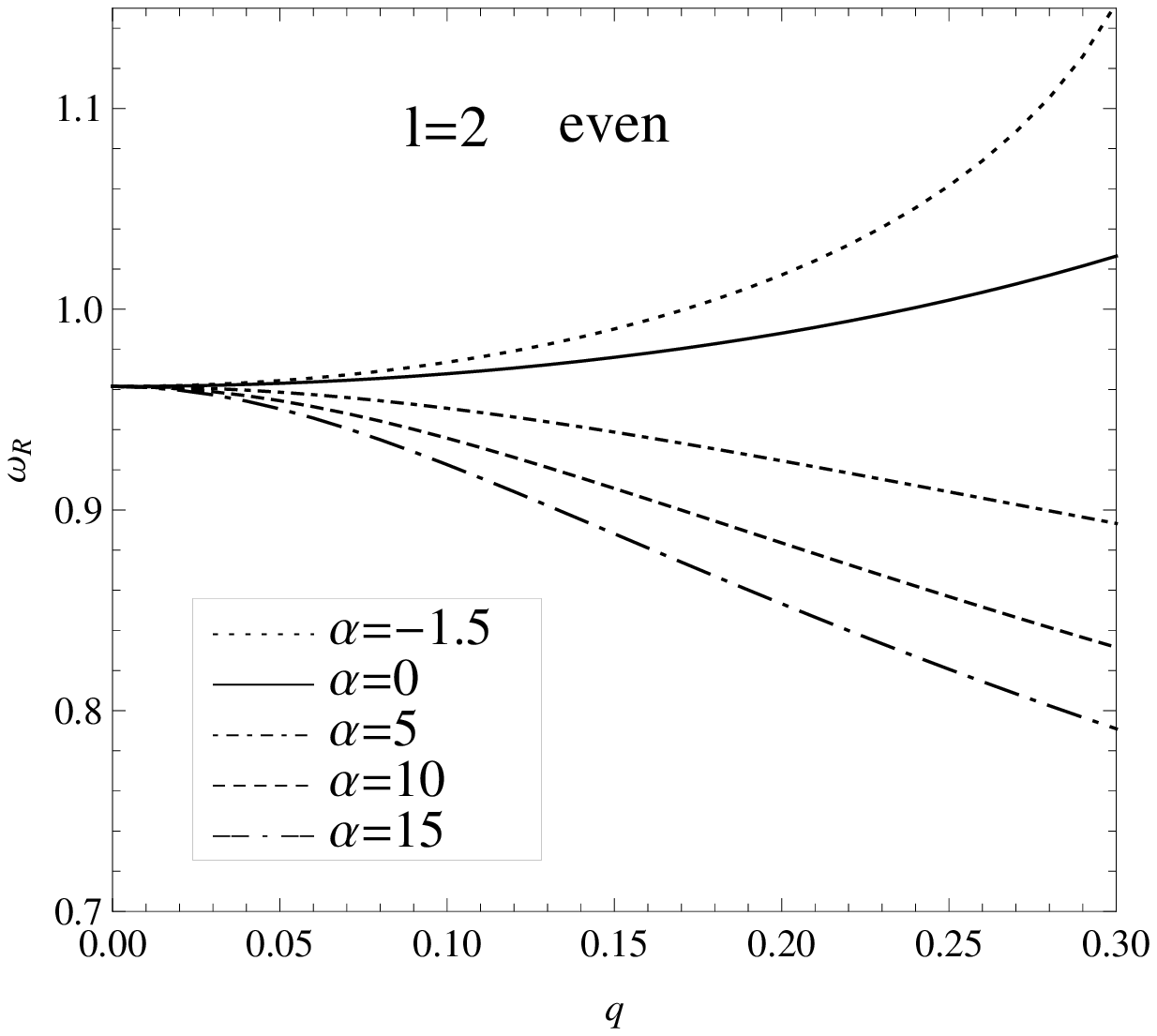}
\includegraphics[width=5.5cm]{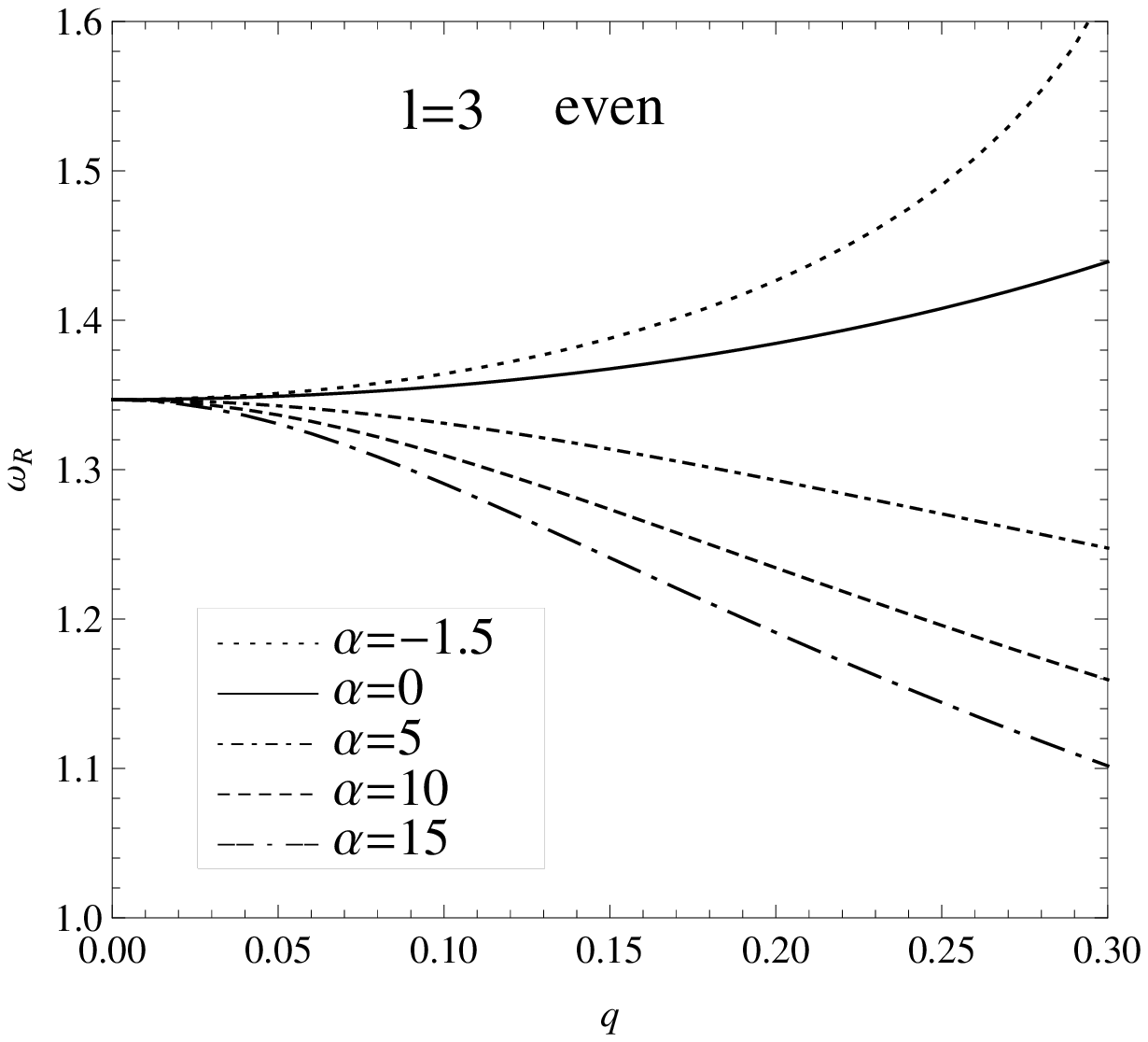}
\caption{Effects of the charge $q$ on the real parts of the fundamental quasinormal modes of vector field perturbation with
the odd parity(the top row) or the even parity (the bottom row)  in
the Reissner-Nordstr\"{o}m black hole spacetime for different $\alpha$. We set $2M=1$.}
\end{center}
\end{figure}
\begin{figure}
\begin{center}
\includegraphics[width=5.5cm]{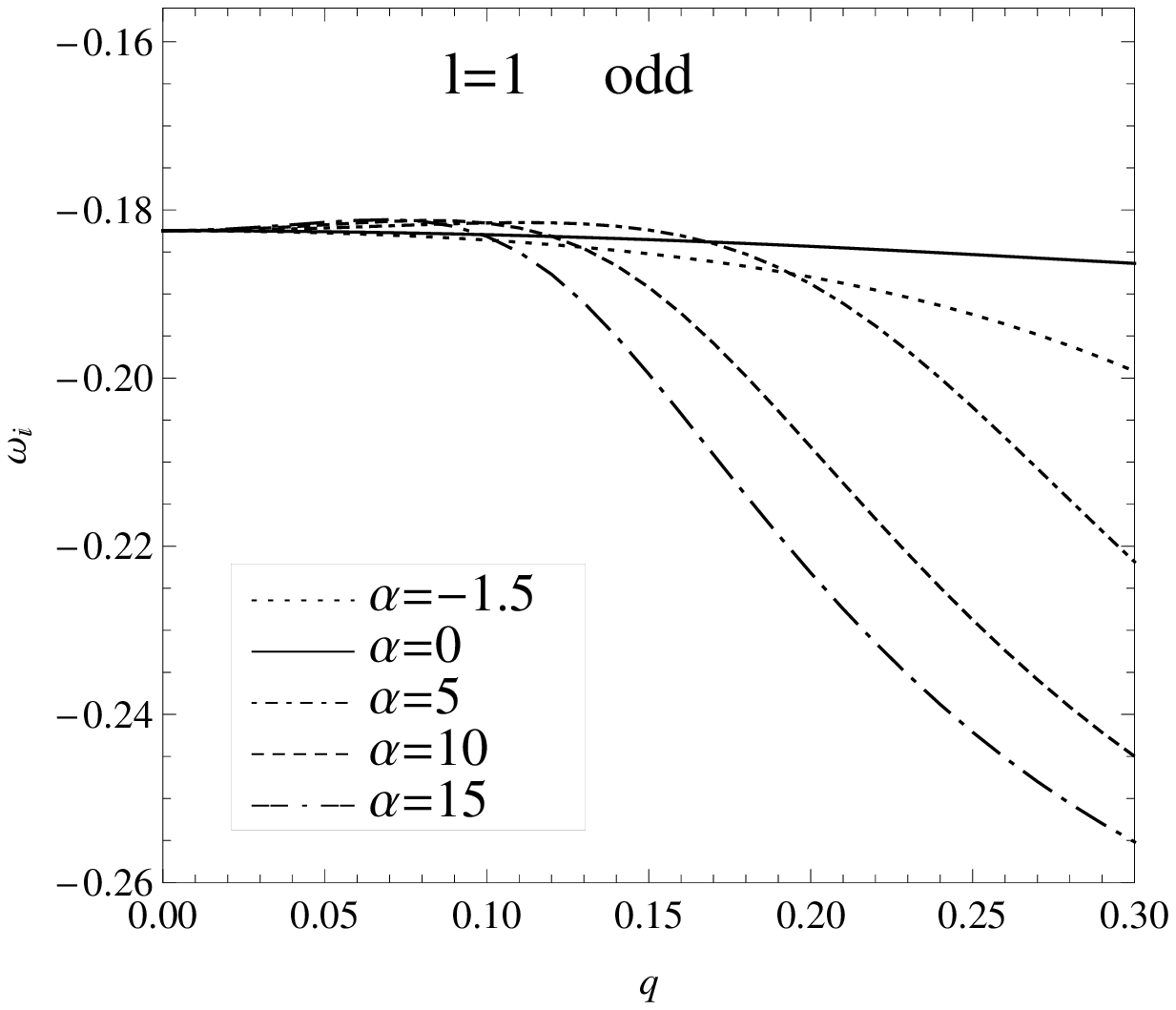}
\includegraphics[width=5.5cm]{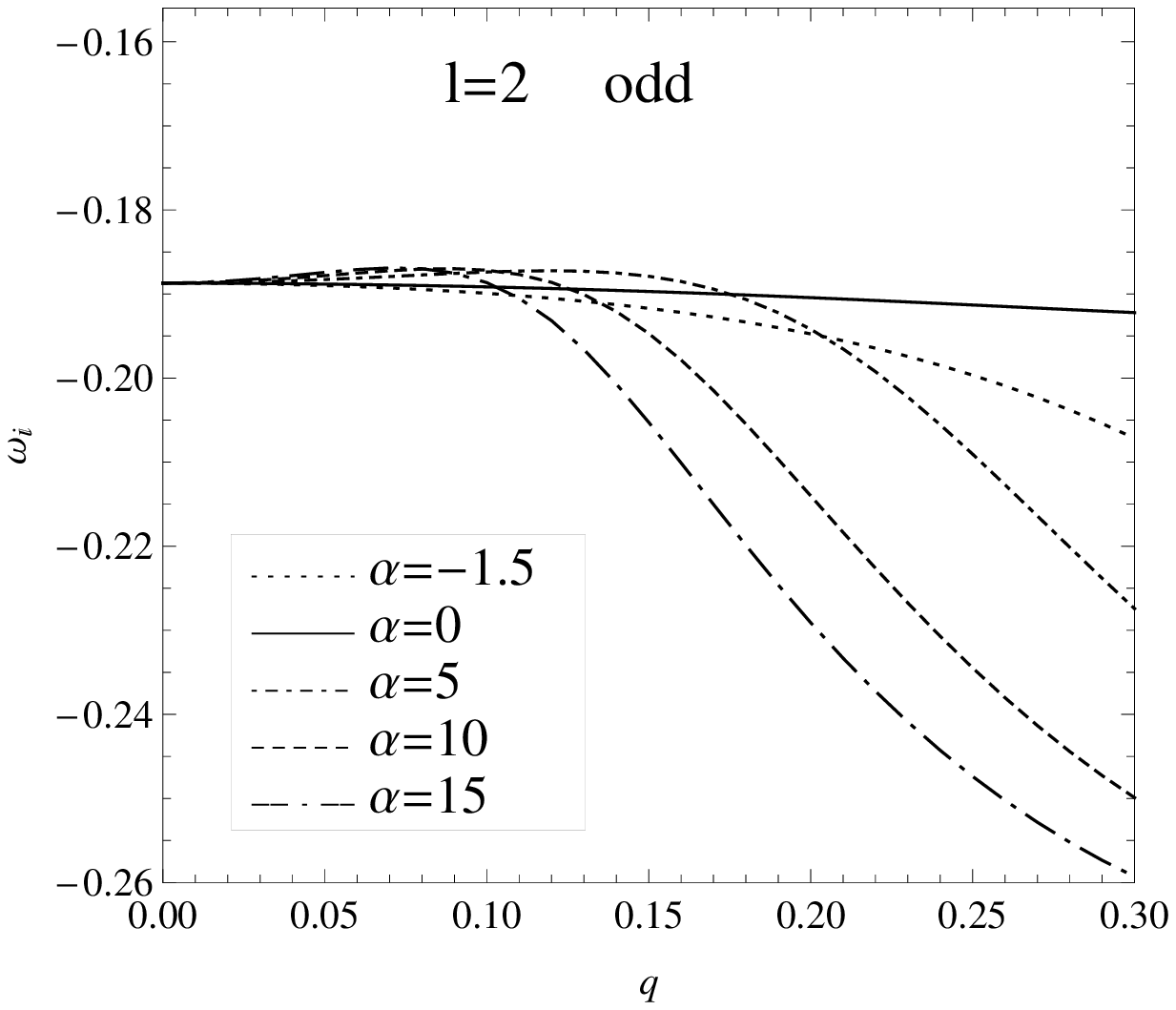}
\includegraphics[width=5.5cm]{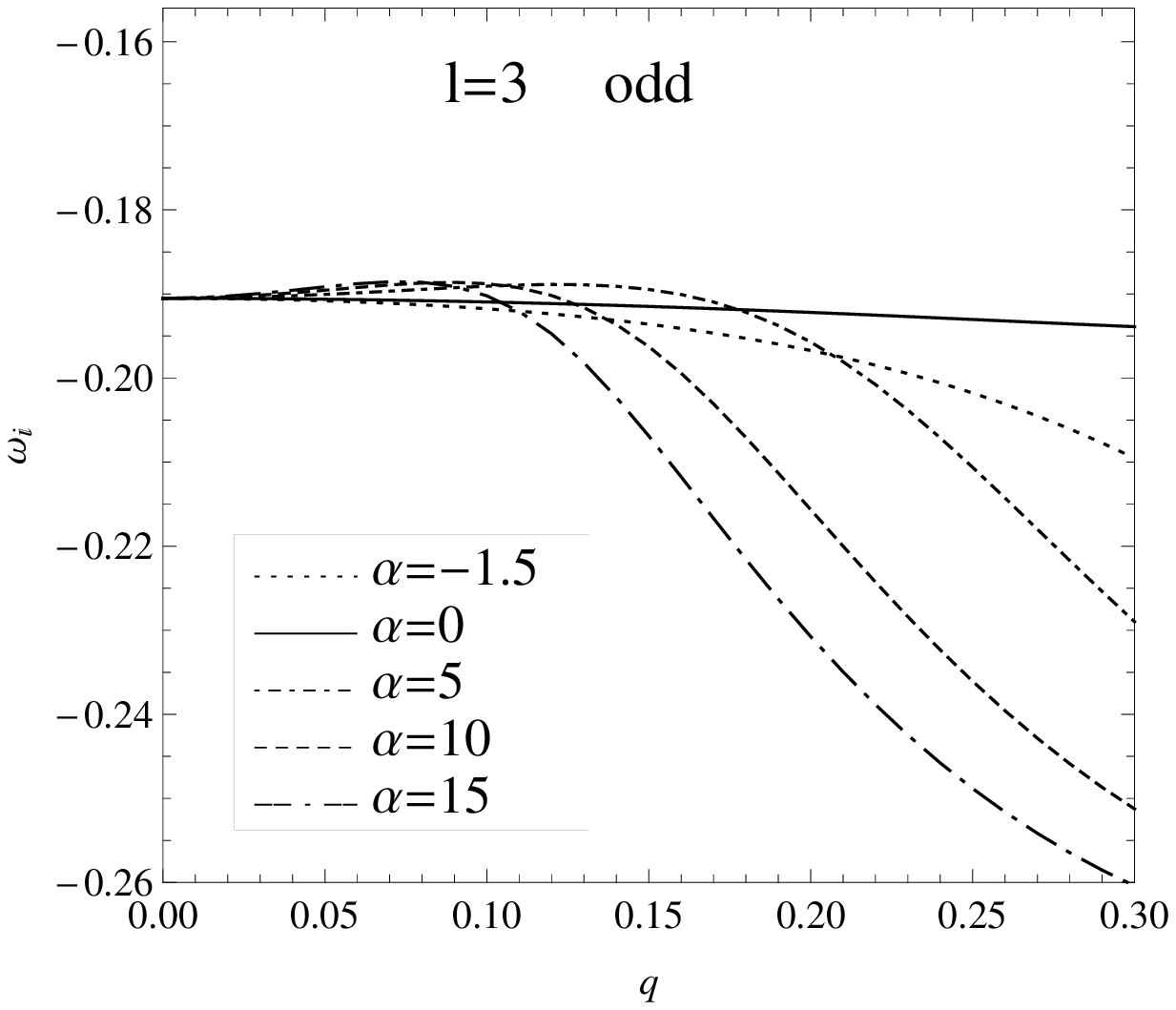}\\
\includegraphics[width=5.5cm]{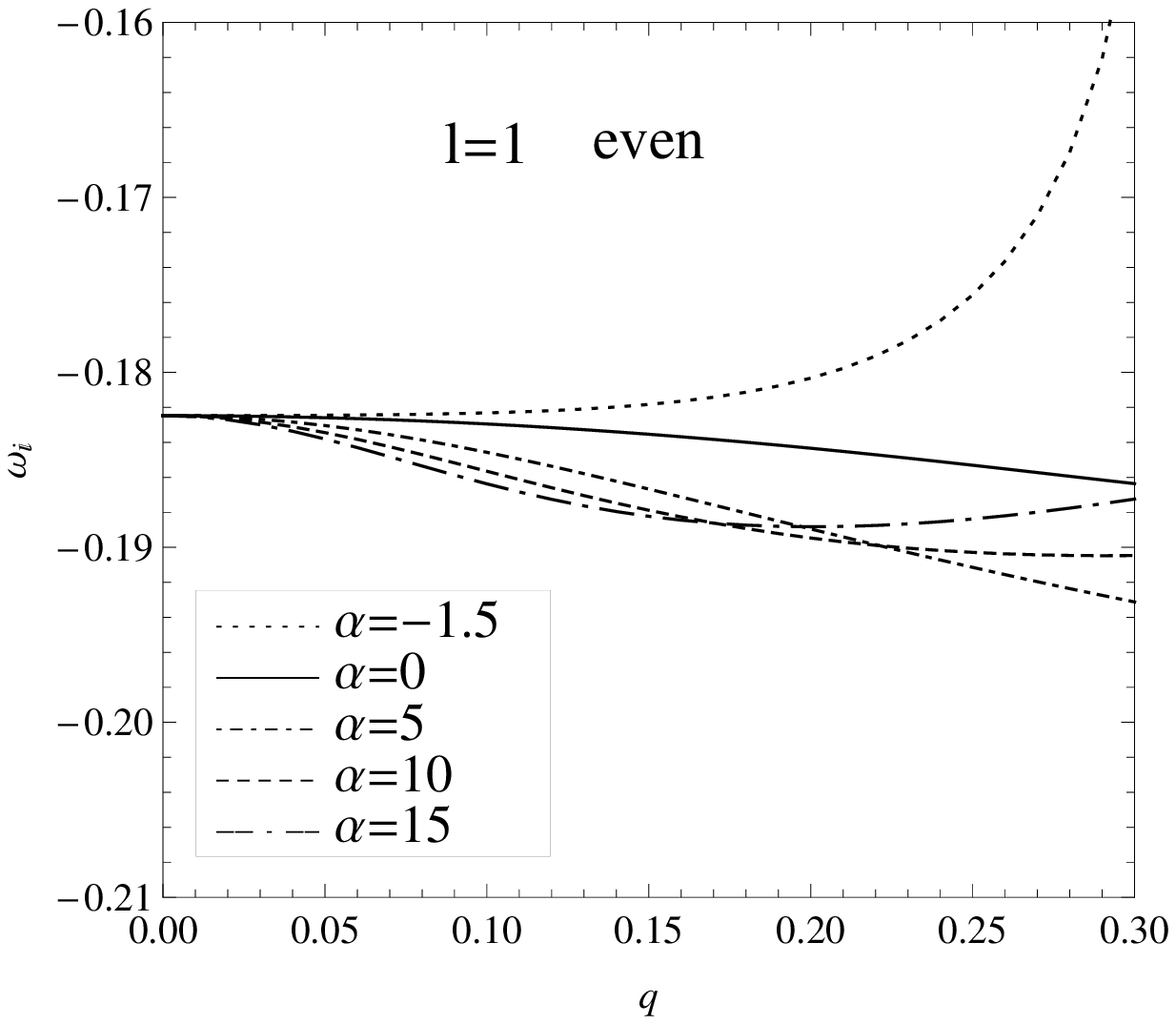}
\includegraphics[width=5.5cm]{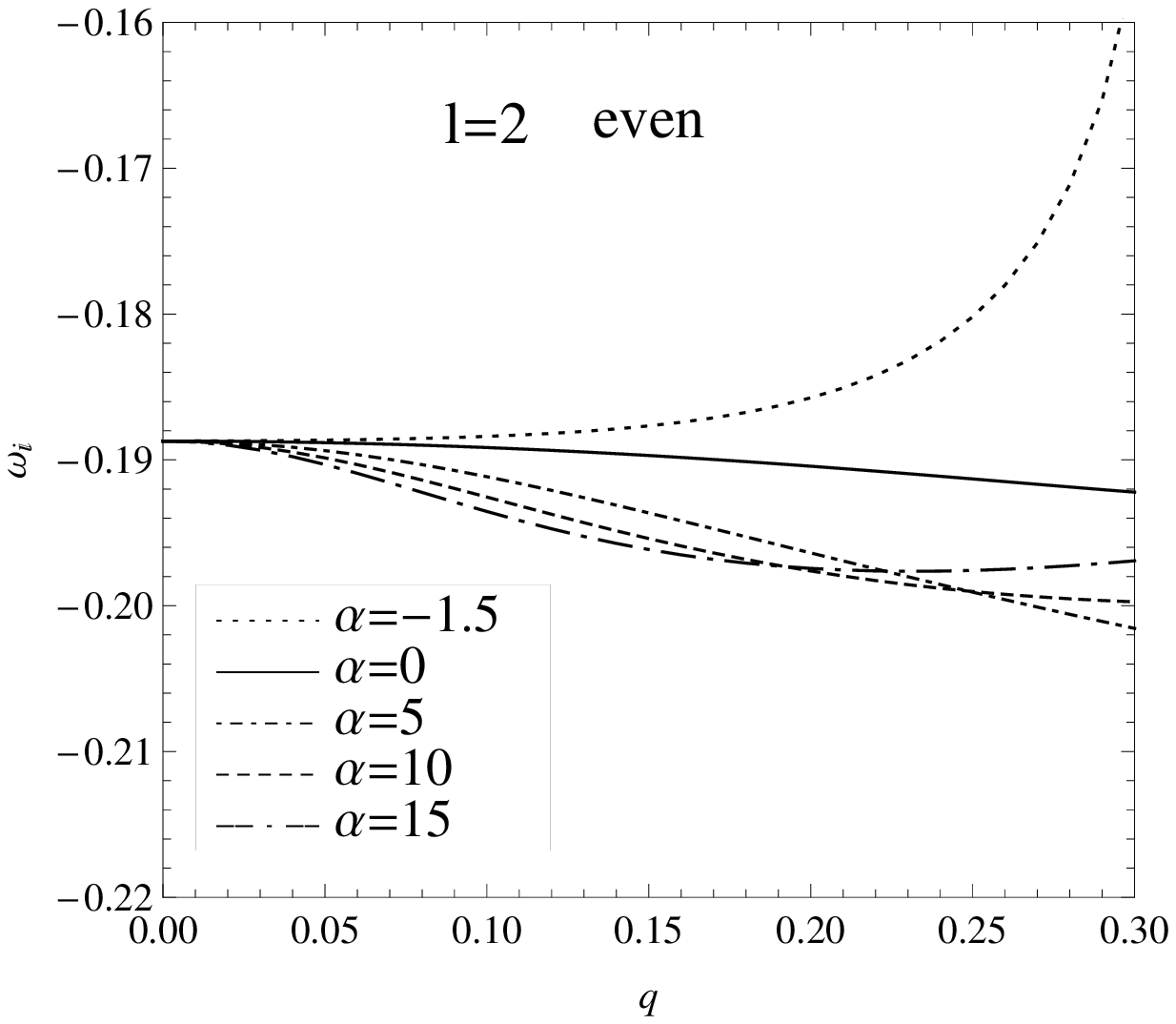}
\includegraphics[width=5.5cm]{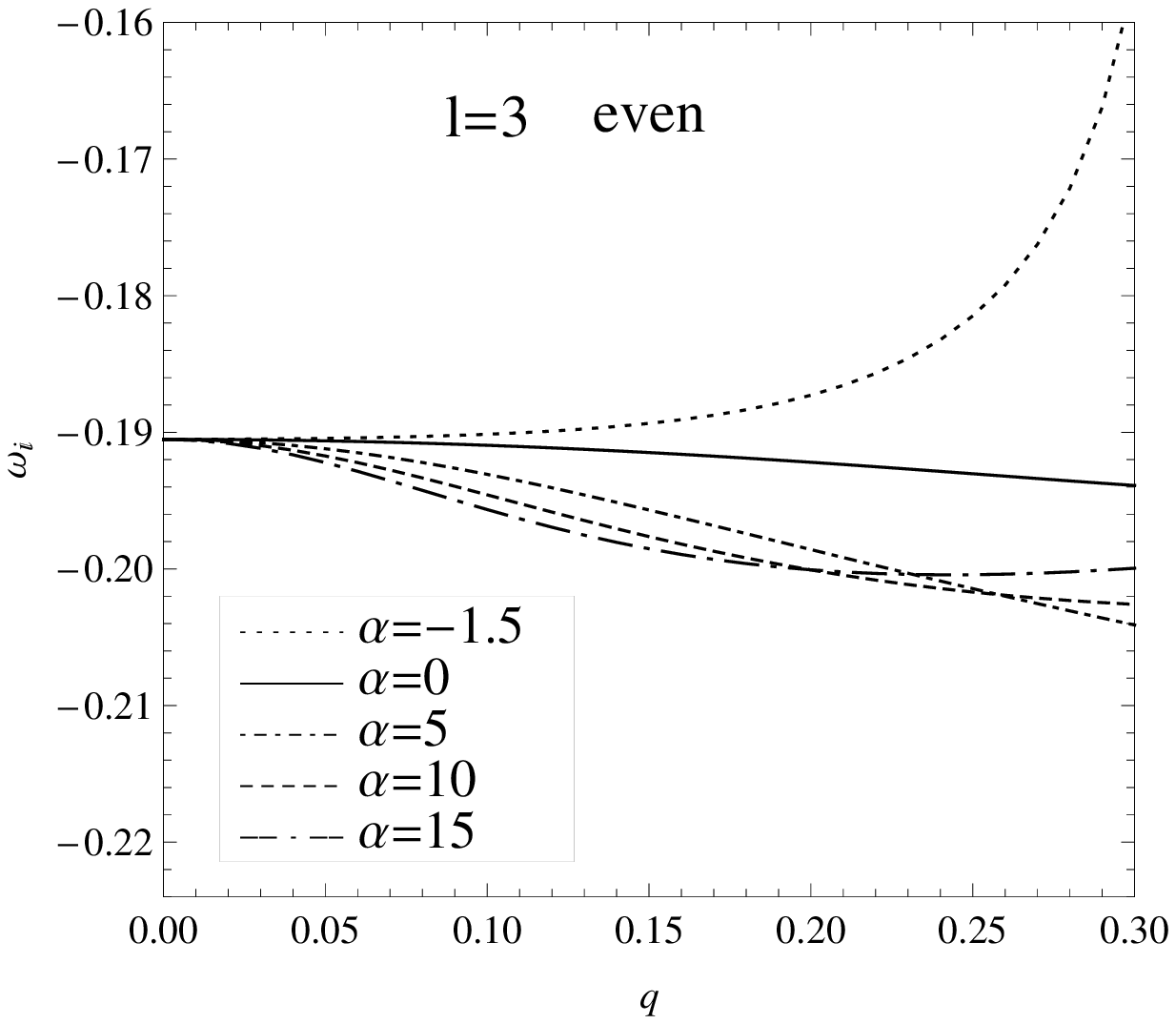}
\caption{Effects of the charge $q$ on the imaginary parts of the fundamental quasinormal modes of vector field perturbation with
the odd parity(the top row) or the even parity (the bottom row)  in
the Reissner-Nordstr\"{o}m black hole spacetime for different $\alpha$. We set $2M=1$.}
\end{center}
\end{figure}
The changes of the imaginary parts with
$\alpha$ are shown in Fig. (4). For the smaller $q$, we find that with increase of $\alpha$  the imaginary parts increase for the odd parity perturbation and decrease for the even parity one. This means
that due to the presence of the coupling parameter the odd parity perturbation decays more slowly and the even parity one decays more quickly in this case.
However, for the larger $q$, we find that the change of the imaginary parts with $\alpha$ exhibits a peak shape for the odd parity vector perturbation. The position of the peak lies in the region $\alpha>0$ and the odd parity perturbation decays most slowly at this position.
For the even parity vector field perturbation in the case with larger $q$,
the imaginary part for the perturbation with the larger $l$ decreases with $\alpha$, but it first decreases and then increases for the perturbation with the smaller $l$.

The dependence of the fundamental quasinormal frequencies on the charge $q$ are plotted in Figs. (5) and (6) for different $\alpha$. The real parts of the
odd parity perturbation increase monotonously with the charge $q$. For the even parity perturbation, the real parts increase monotonously with $q$ in the case of $\alpha<0$ and decrease monotonously with $q$ in the case of $\alpha>0$. From Fig.(6), one can find that for the negative $\alpha$, the imaginary part decreases with $q$ for the odd parity perturbation and increases for the even parity perturbation. For the positive $\alpha$, the imaginary parts of the odd parity perturbation first increases and then decreases with $q$. The changes of imaginary part of the even parity perturbation is just the opposite to that of the odd parity perturbation in this case. Thus, both of the oscillation and the decay rate of vector perturbations in the  Reissner-Nordstr\"{o}m
black hole spacetime depend on the coupling between the perturbational field and Einstein tensor, the black hole parameters, the multipole number $l$ and parity of perturbational field itself.

In the subsequent section, we will study the dynamical evolution of the
vector field perturbation coupling to Einstein tensor in time domain \cite{Gundlach:1994} and examine if the coupled perturbation is stable or not in the background of a Reissner-Nordstr\"{o}m black hole. With the help of the light-cone variables $u=t-r_*$ and $v=t+r_*$, one can find
that the wave equation
\begin{eqnarray}
-\frac{\partial^2\psi}{\partial t^2}+\frac{\partial^2\psi}{\partial
r_*^2}=V(r)\psi,
\end{eqnarray}
can be rewritten as
 \begin{eqnarray}
4\frac{\partial^2\psi}{\partial u\partial
v}+V(r)\psi=0.\label{wbes1}
\end{eqnarray}
This two-dimensional wave equation (\ref{wbes1}) can be integrated numerically by using the finite difference method \cite{Gundlach:1994}. Making use of Taylor's theorem, one can find that the wave equation (\ref{wbes1}) can be
discretized as
\begin{eqnarray}
\psi_N=\psi_E+\psi_W-\psi_S-\delta u\delta v
V(\frac{v_N+v_W-u_N-u_E}{4})\frac{\psi_W+\psi_E}{8}+O(\epsilon^4)=0.\label{wbes2}
\end{eqnarray}
Here the points $N$, $W$, $E$ and $S$ are defined as: $N$:
$(u+\delta u, v+\delta v)$, $W$: $(u + \delta u, v)$, $E$: $(u, v +
\delta v)$ and $S$: $(u, v)$. The parameter $\epsilon$ is an overall
grid scalar factor, so that $\delta u\sim\delta v\sim\epsilon$.
As in \cite{Gundlach:1994}, we can set $\psi(u, v=v_0)=0$ and use a
Gaussian pulse as an initial perturbation, centered on $v_c$ and
with width $\sigma$ on $u=u_0$ as
\begin{eqnarray}
\psi(u=u_0,v)=e^{-\frac{(v-v_c)^2}{2\sigma^2}}.\label{gauss}
\end{eqnarray}
\begin{figure}[ht]
\begin{center}
\includegraphics[width=5.5cm]{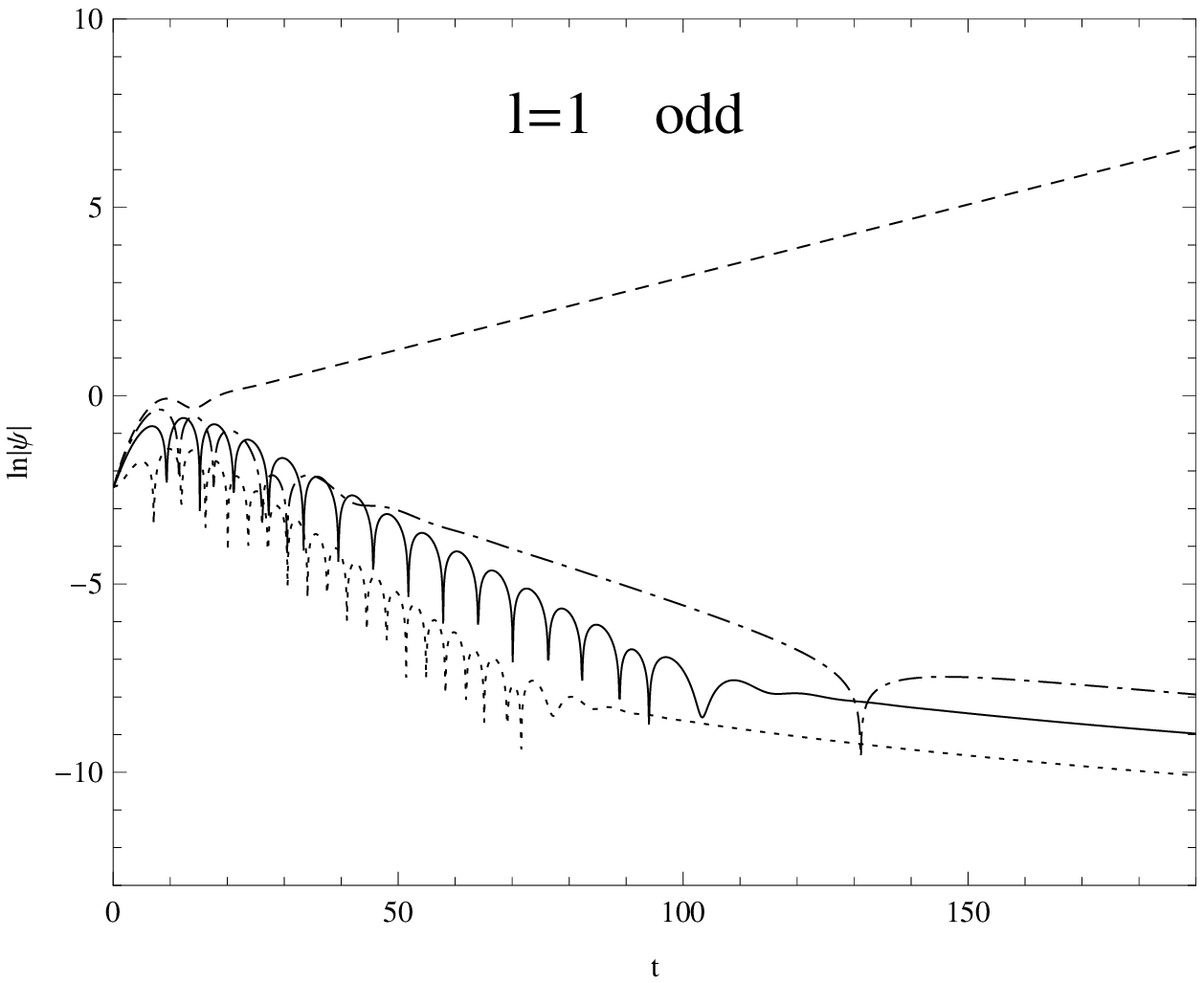}\includegraphics[width=5.5cm]{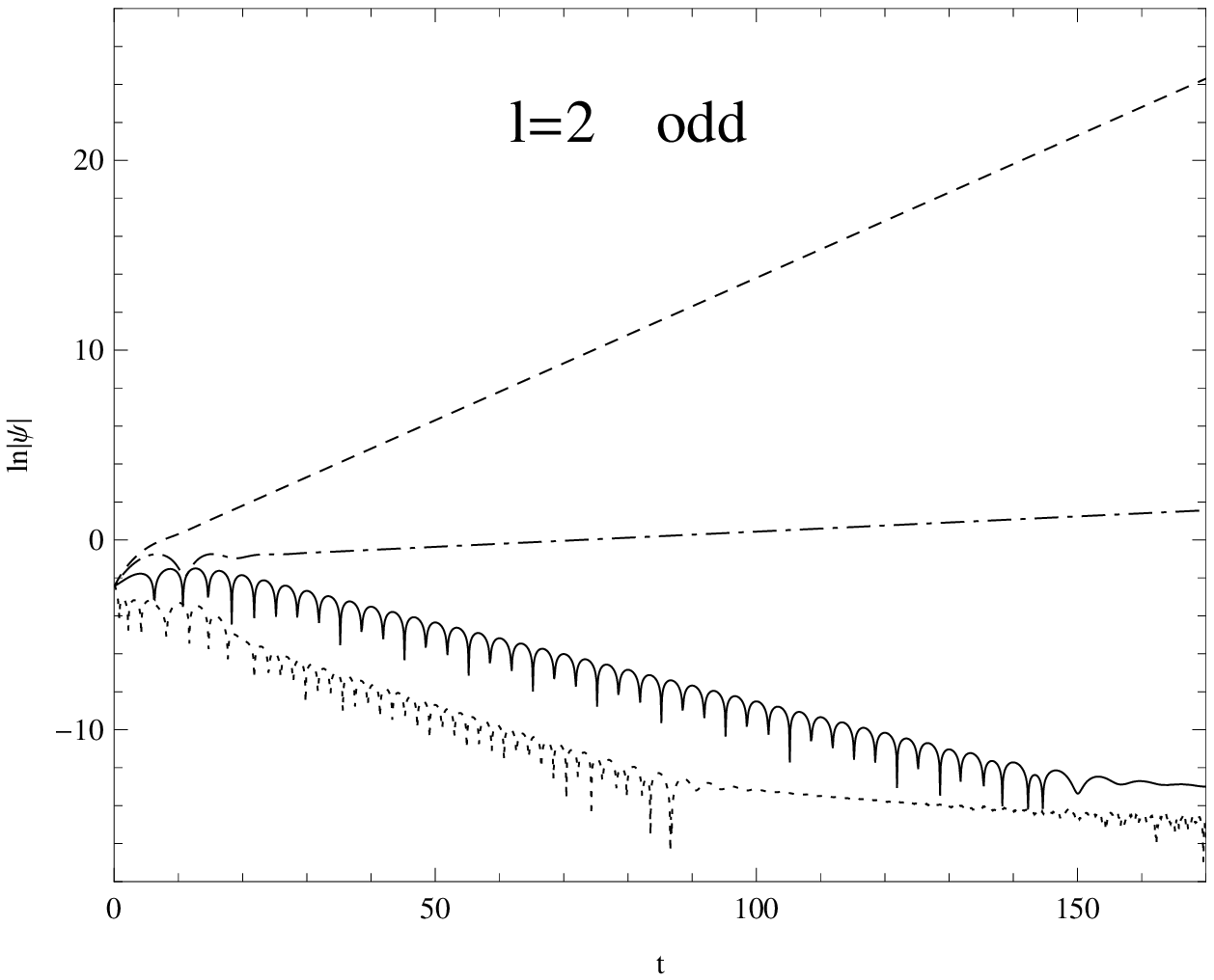}
\includegraphics[width=5.5cm]{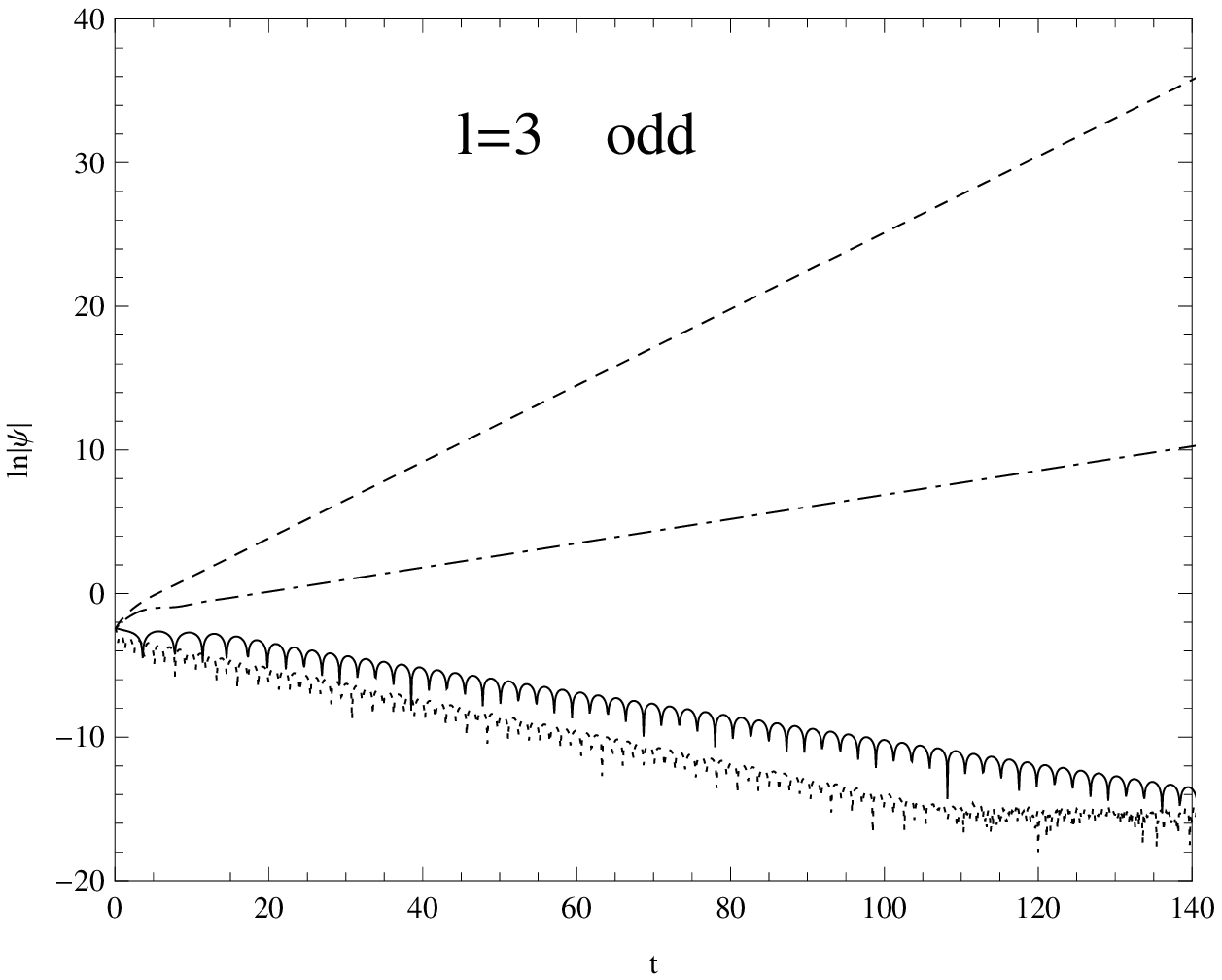}
\caption{The dynamical evolution of an vector field perturbation with odd parity in the background of a Reissner-Nordstr\"{o}m black hole spacetime. The figures from left to right are corresponding to $l=1$, $2$ and $3$. The dotted, solid, dash-dotted and dashed lines are corresponding to the cases with $\alpha=30,~0,~-10,~-15$, respectively. We set $2M=1$ and $q=0.2$. The constants in the Gauss pulse (\ref{gauss}) $v_c=10$ and $\sigma=3$.}
\end{center}
\end{figure}
\begin{figure}[ht]
\begin{center}
\includegraphics[width=5.5cm]{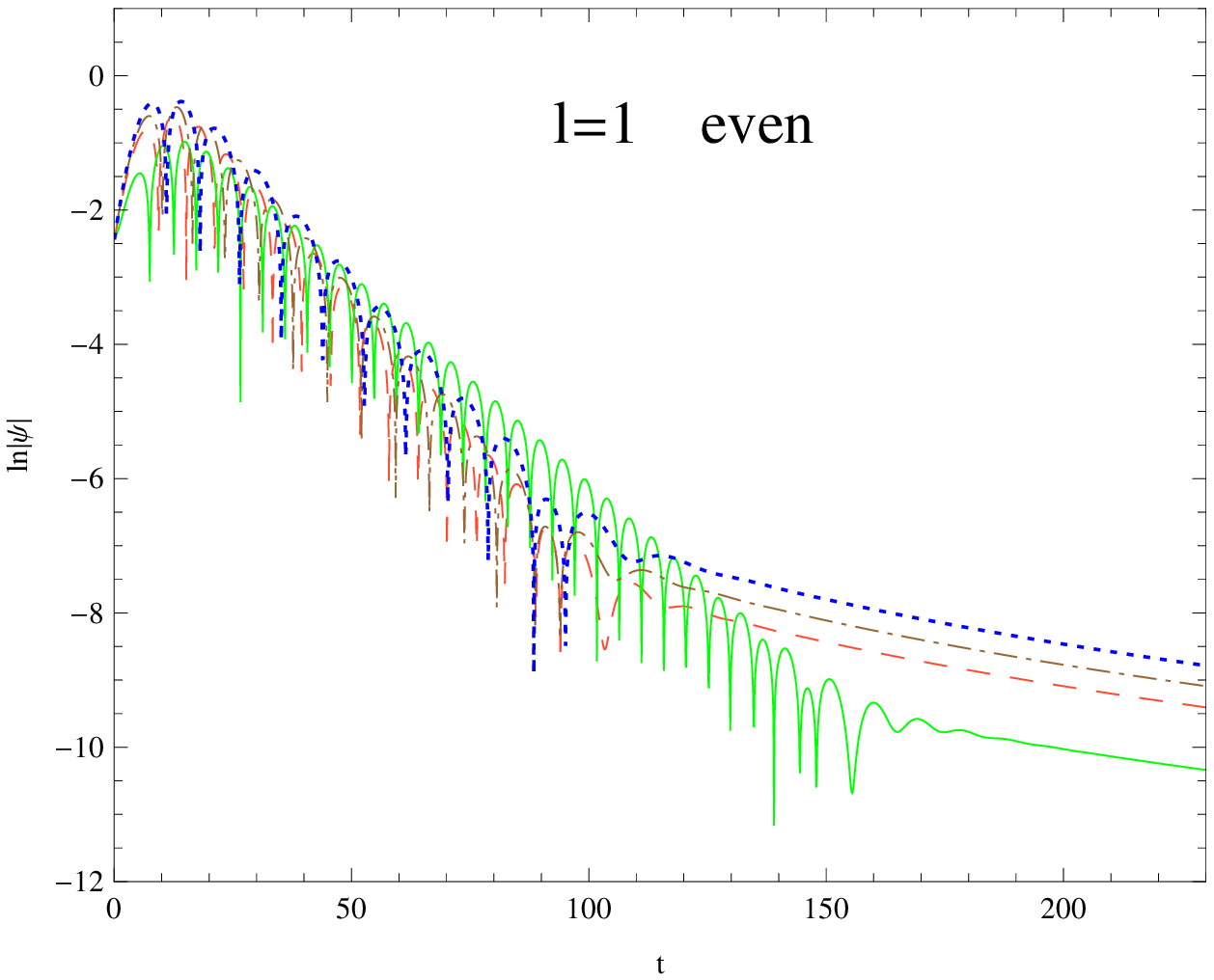}\includegraphics[width=5.5cm]{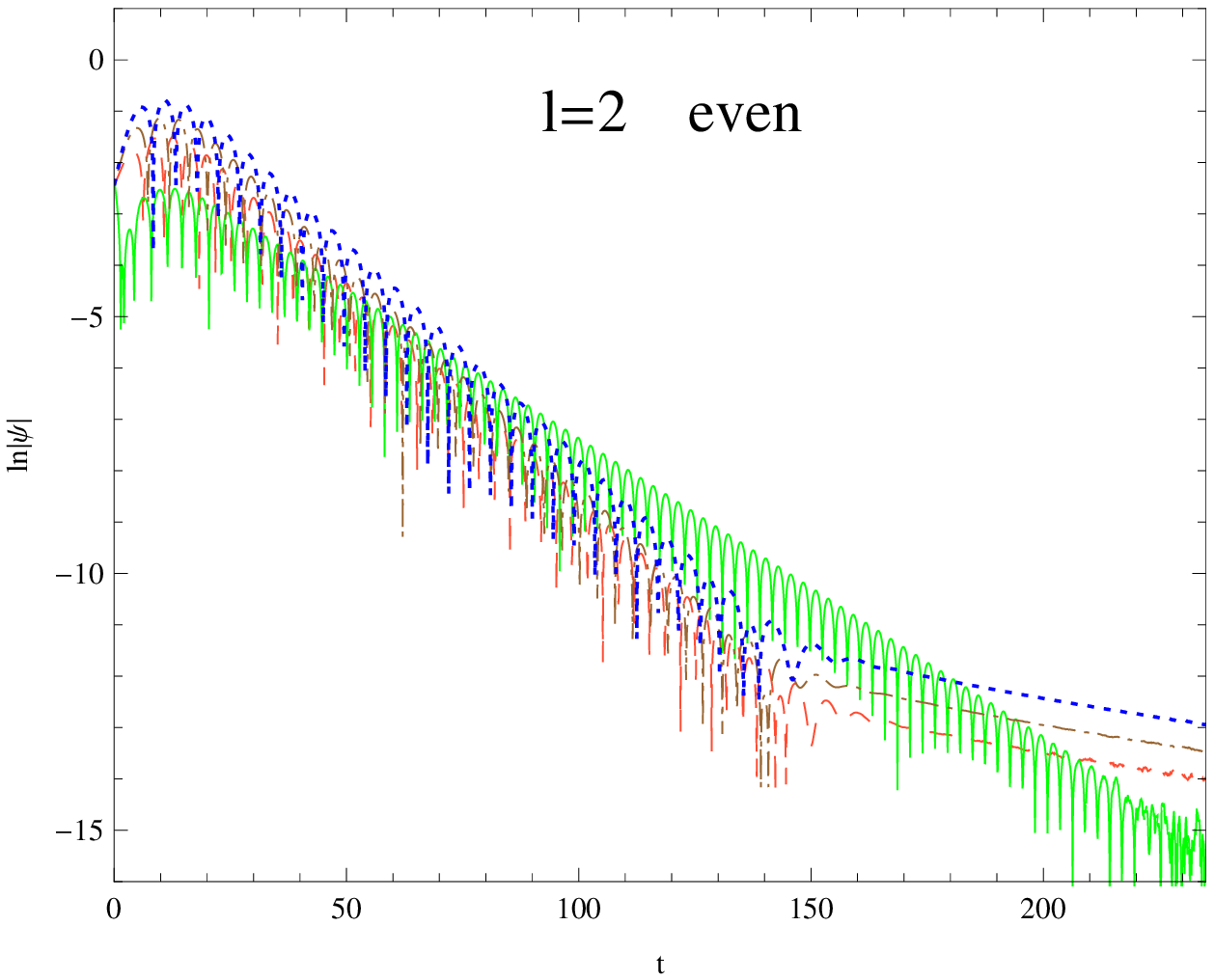}
\includegraphics[width=5.5cm]{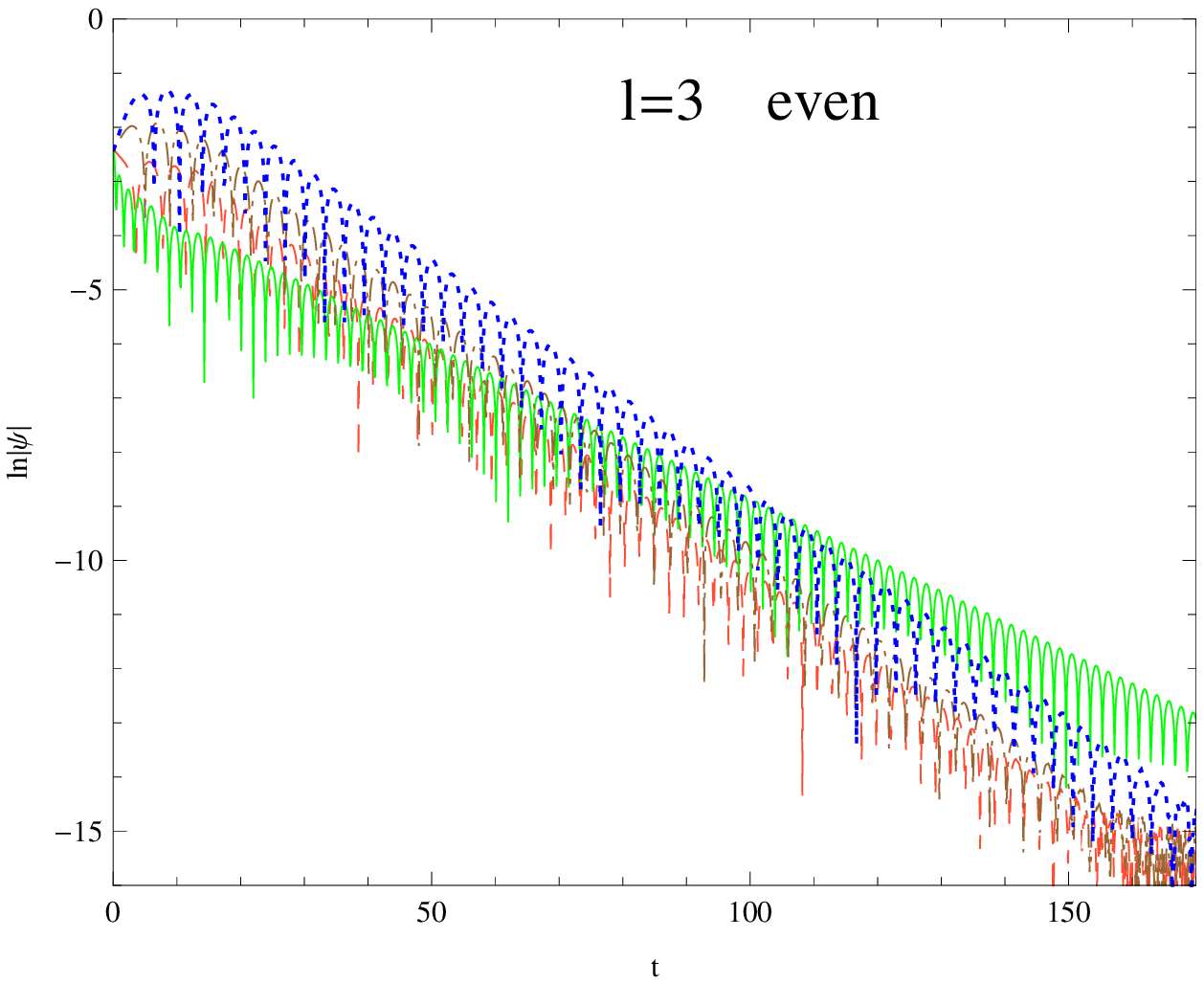}
\caption{The dynamical evolution of an vector field perturbation with even parity in the Reissner-Nordstr\"{o}m  black hole spacetime. The figures from left to right are corresponding to $l=1$, $2$ and $3$. The solid, dashed, dash-dotted and dotted lines are corresponding to the cases with $\alpha=-5,~0,~10,~40$, respectively. We set $2M=1$ and $q=0.2$. The constants in the Gauss pulse (\ref{gauss}) $v_c=10$ and $\sigma=3$.}
\end{center}
\end{figure}
In figs. (7) and (8), we show the dynamical evolution of the external vector field perturbation coupling to Einstein tensor in the
Reissner-Nordstr\"{o}m black hole spacetime, which indicates that the dynamical evolution of the coupled vector field perturbation depends heavily on the parity of the perturbation field.

For the  perturbational field with even parity, it always decays for arbitrary multipole number $l$ and the coupling constant $\alpha$.
For the positive $\alpha$, we find that the odd parity perturbation also decays as the usual perturbation without coupling to Einstein tensor in the black hole spacetime. However, for the negative $\alpha$, we find that if the coupling constant $\alpha$ is below the critical value $\alpha_c$, the odd parity vector field does not decay any longer, and it grows with exponential rate, which means that
the unstable mode of the vector field perturbation appears as in \cite{RAK1,RAK2,Card}. We must point out that the final stage of the evolution of a growing perturbation can not be described by the perturbational theory because initially small perturbations will become comparable with the background at sufficiently late times and the previous perturbational method is not valid again. The perturbational method provides us only an initial prediction for the developing of the instability. Moreover, we also find that the stronger coupling leads to that the unstable mode appears more early and the growth rate becomes larger. This could be explained by a fact that as $\alpha<0$, the stronger coupling drops down the peak of the potential barrier and increases the negative gap near the black hole horizon for the odd parity vector perturbation, which could result in that the effective potential is non-positive definite.
For the vector field perturbation with even parity, it always decays in the allowed range of $\alpha$. The main reason is that in its effective potential there does not exist the negative gap near the black hole horizon. These dynamical properties of vector field perturbation coupling to Einstein tensor are very similar to those of electromagnetic field with Weyl corrections \cite{sb2013}.

For the odd parity vector field perturbation coupling to Einstein tensor, the dependence of the critical value $\alpha_c$
on the multipole number $l$ is plotted in Fig.(9), which shows that the critical value $\alpha_c$ can be fitted best by the function
\begin{eqnarray}
\alpha_c\simeq a\sqrt{\frac{l}{l+1}}+b,\label{ncs}
\end{eqnarray}
where the coefficients $a$ and $b$ are numerical constants with dimensions of length-squared and their values depend on the charge $q$ of black hole.
\begin{figure}
\begin{center}
\includegraphics[width=5.5cm]{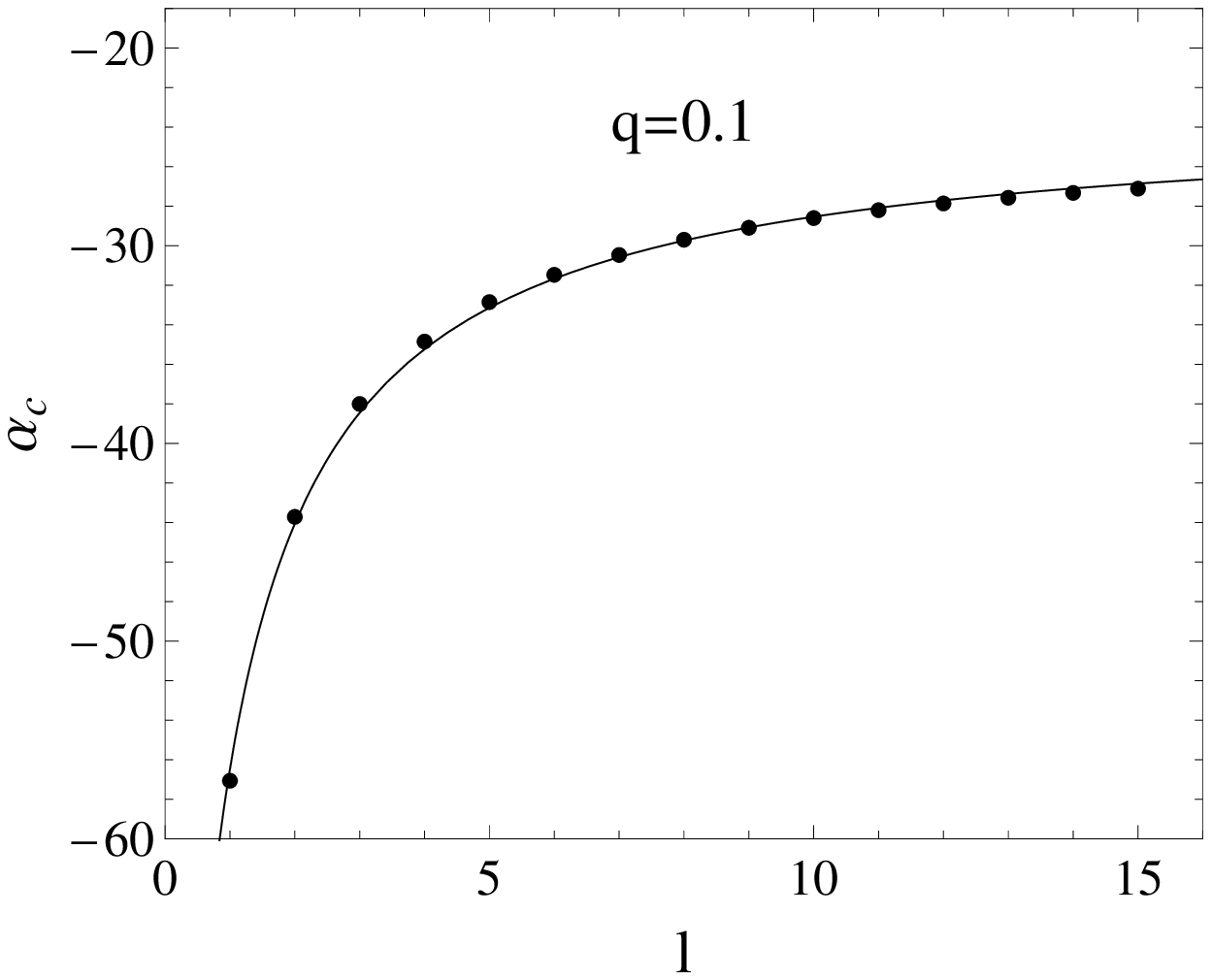}\includegraphics[width=5.5cm]{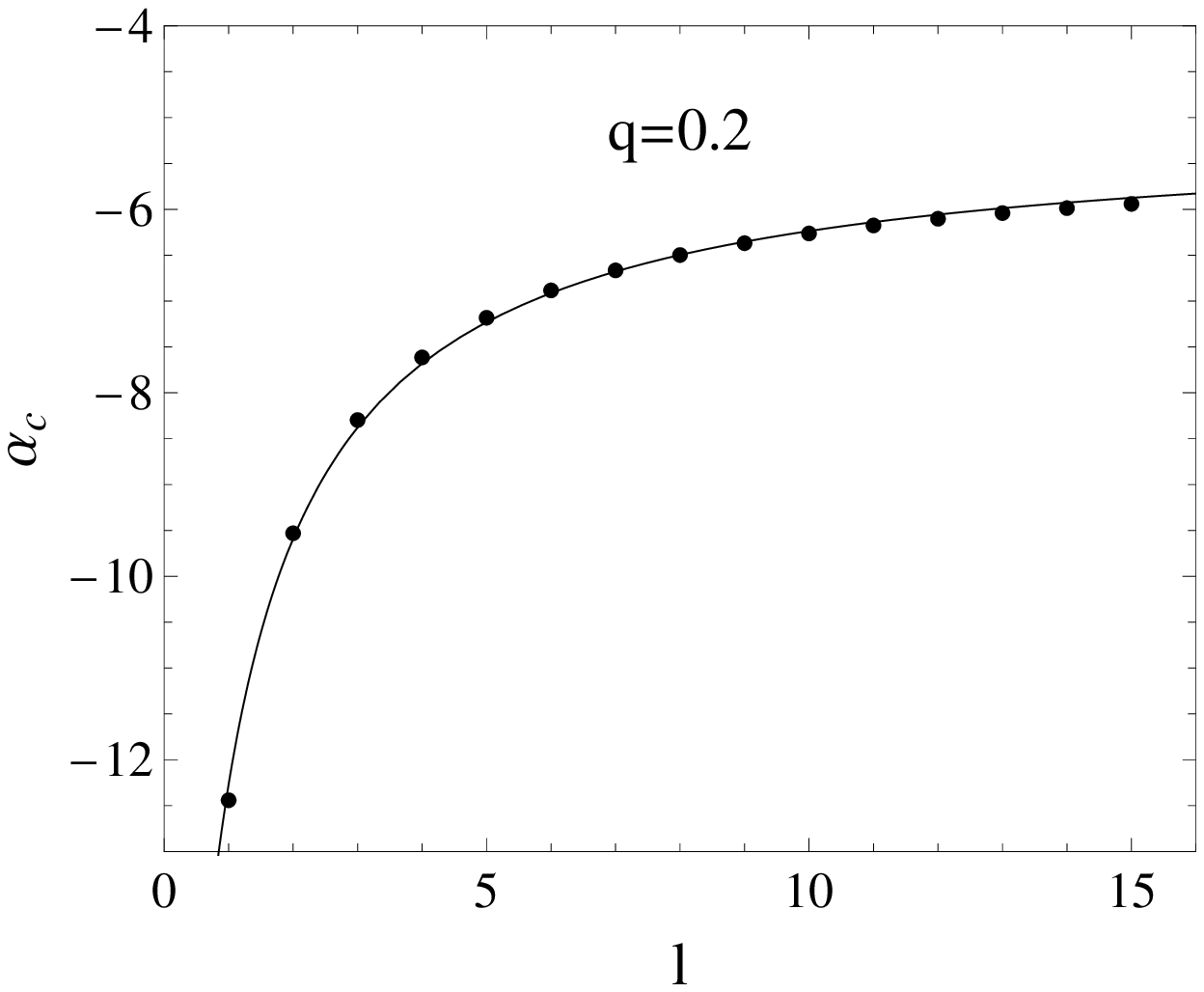}
\includegraphics[width=5.5cm]{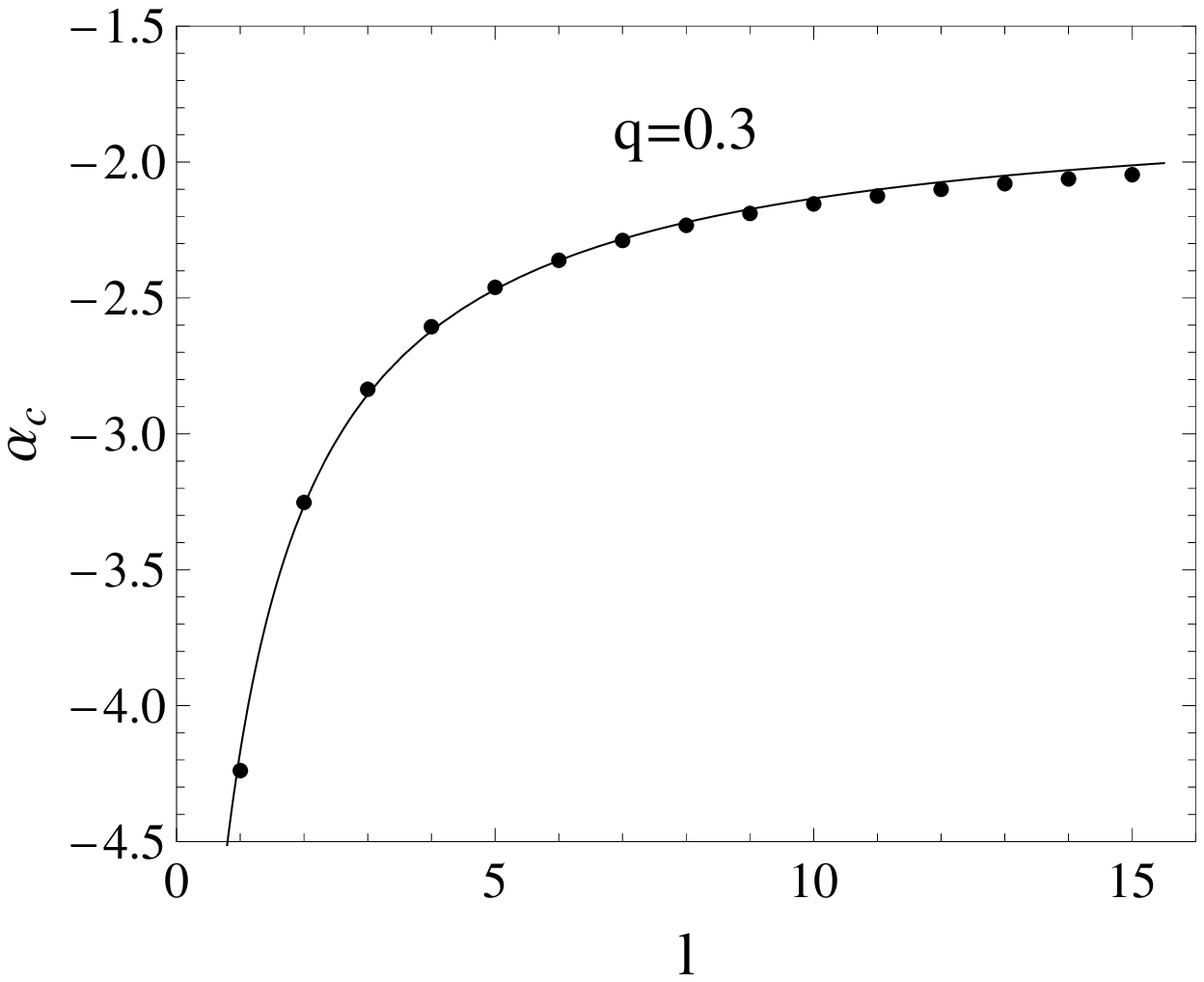}
\caption{The change of the threshold value $\alpha_c$ with the multipole number $l$ for fixed $q$. The points $l=1\sim 15$ are fitted by the function $\alpha_c=a~\sqrt{\frac{l}{l+1}}+b$, where the coefficients $a$ and $b$ are numerical constants with dimensions of length-squared and their values depend on the charge $q$ of black hole.}
\end{center}
\end{figure}
Obviously, one can find that the threshold value $\alpha_c$ is negative and the absolute value $|\alpha_c|$ decreases with the multipole number $l$ and $q$, which means that for the higher $l$ and $q$ the unstable mode appears more easily. The changes of the numerical constants $a$ and $b$ with the charge
$q$ are presented in Fig. (10), which tells us that
the values of $a$, $b$ and $a+b$ are fitted by the functions $a\simeq -4.1728 +\frac{1.2002}{q^2}-\frac{0.2436}{q}$, $b\simeq 4.9811 -\frac{1.4454}{q^2}+ \frac{0.2890}{q}$ and $a+b\simeq -\frac{r^4_+}{4q^2}$,
respectively. Therefore, for a Schwarzschild black hole, one can find that the threshold value $\alpha_c$ tends to infinity for arbitrary $l$ since the charge $q$ disappears, which implies that for the vector field coupling to Einstein¡¯s tensor, it always decays and the unstable mode does not appear in the Schwarzschild black hole spacetime. It is not surprising because for a Schwarzschild black hole all the components of the Einstein tensor vanish.
\begin{figure}
\begin{center}
\includegraphics[width=5.5cm]{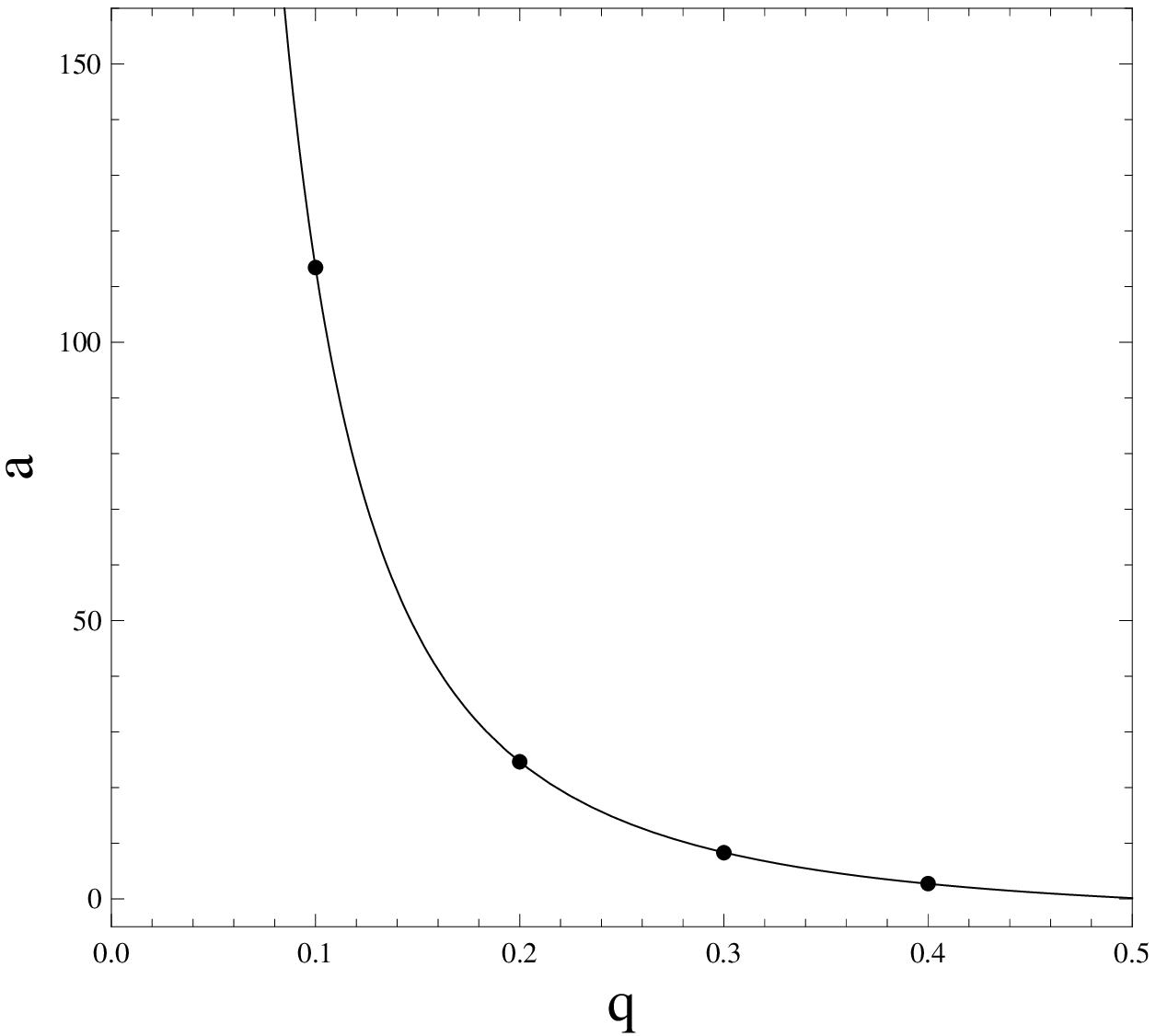}\includegraphics[width=5.5cm]{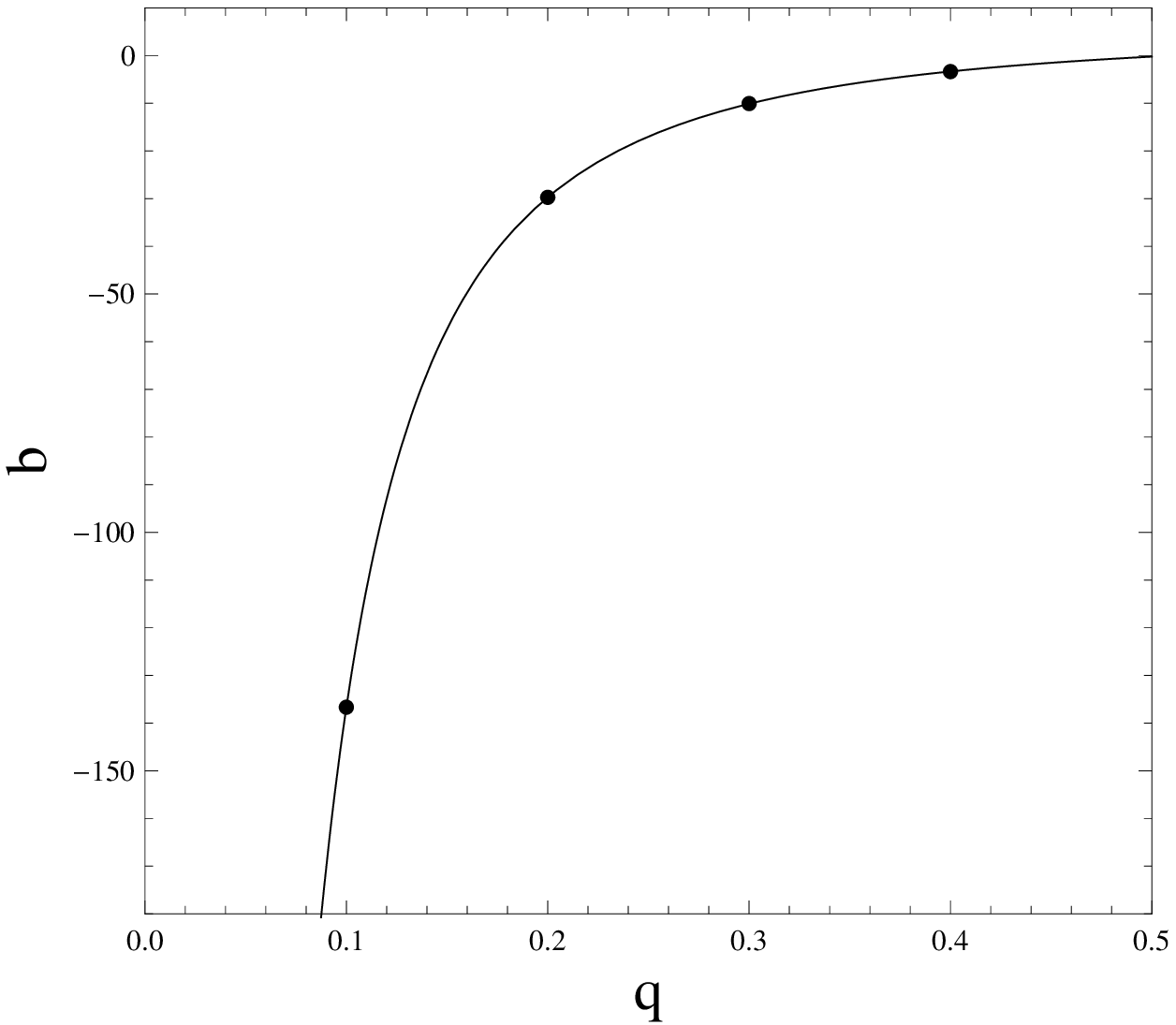}
\includegraphics[width=5.5cm]{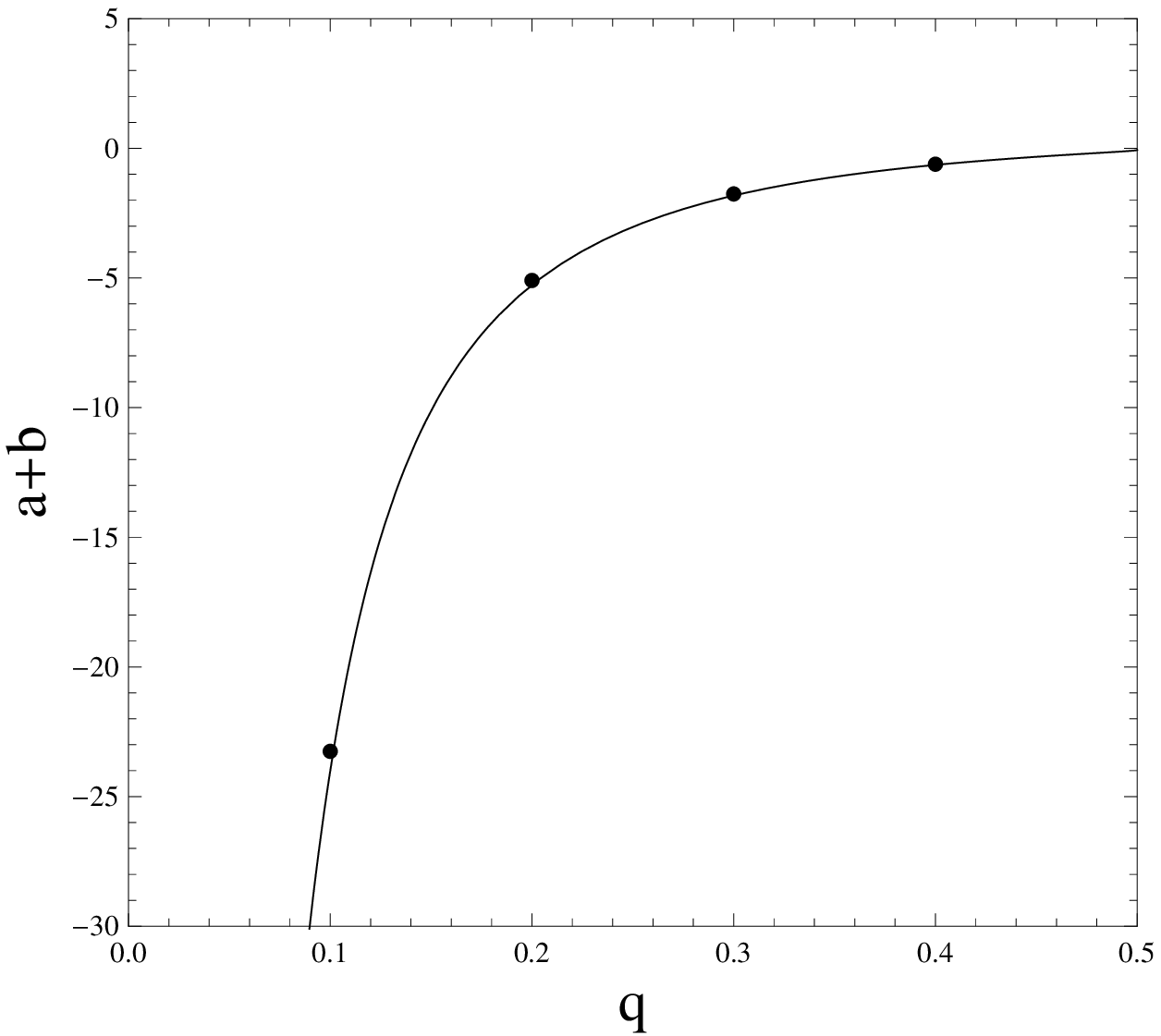}
\caption{The changes of the numerical constants $a$, $b$ and $a+b$ with the charge $q$, which are fitted by the functions $a\simeq -4.1728 +\frac{1.2002}{q^2}-\frac{0.2436}{q}$, $b\simeq 4.9811 -\frac{1.4454}{q^2}+ \frac{0.2890}{q}$  and $a+b\simeq -\frac{r^4_+}{4\; q^2}$,
respectively.}
\end{center}
\end{figure}
Moreover, from Eq. (\ref{ncs}) and Fig.(10), we can obtain that as $l\rightarrow\infty$
the threshold value $\alpha_c\rightarrow a+b\simeq\frac{r^4_+}{4\; q^2}$. The main reason is that in this limit the integration outside the outer event horizon \cite{Gleiser:2005},
\begin{eqnarray}
&&(\Psi(r),\omega^2\Psi(r))=\int^{\infty}_{-\infty}\Psi^{*}(r)\bigg(-\frac{d^2}{dr^2_*}
+V(r)\bigg)\Psi(r)dr_*\bigg|_{l\rightarrow\infty}
\simeq\int^{\infty}_{r_+}|\Psi(r)|^2\frac{V(r)}{f}dr
\nonumber\\
&&=\int^{\infty}_{r_+}|\Psi(r)|^2\bigg(\frac{r^4+4\alpha
q^2}{r^4}\bigg)\frac{l(l+1)}{r^2}dr,
\end{eqnarray}
is positive definite as $\alpha<-r^4_+/(4q^2)$, which implies that the
threshold value $\alpha_c=-r^4_+/(4q^2)$ as $l\rightarrow\infty$.

Finally, we make a comparison among the dynamical properties of perturbational fields coupling to gravitational tensors in some black hole spacetimes. For the scalar field coupling to Einstein tensor in the Reissner-Nordstr\"{o}m  black hole spacetime \cite{sb2010}, we find that the partial wave $l=0$ always decays for arbitrary coupling constant and the nonzero-$l$ partial waves grow with exponential rate as the coupling constant $\eta$ is larger than the certain a critical $\eta_c$. For the electromagnetic field coupling to Weyl tensor \cite{sb2013} or the vector field coupling to Einstein tensor, we find that the dynamical properties of perturbations depend sharply on the parities of perturbational fields themselves and the coupling strength between perturbational fields and gravitational tensors. For the  perturbational field with even parity, it always decays for arbitrary multipole number $l$ and coupling constant $\alpha$. For the perturbational field  with odd parity grows with exponential rate if the coupling constant $\alpha$ is smaller than certain a negative critical value $\alpha_c$. These dynamical properties of the odd parity vector field coupling to Einstein tensor or Weyl tensor are similar to those of the nonzero-$l$ partial waves of a scalar field coupling to Einstein tensor in the Reissner-Nordstr\"{o}m  black hole spacetime. Moreover, we find also that the critical coupling constant and the change of quasinormal frequencies with the coupling constant depend on the types of the coupled gravitational tensors, the black hole parameters, the multipole number $l$ and spin $s$ of perturbational fields. Finally, our results show that the constraint on the coupling constant $\alpha$ depends on the parity of vector field perturbation as it couples to Einstein tensor or Weyl tensor.

\section{summary}

We have studied the wave dynamics of an vector field perturbation coupling to Einstein tensor in the four-dimensional Reissner-Nordstr\"{o}m black hole spacetime. We find that the wave dynamic equation of the vector field perturbation strongly depends on the parity of the perturbational field itself and the coupling strength between vector field and Einstein tensor. With the increase of the coupling constant $\alpha$, we find that the real parts of fundamental quasinormal frequencies increase monotonously for the vector field perturbation with odd parity and decrease for the perturbation with even parity. The changes of the imaginary parts with
$\alpha$ are more complicated, which depend on the coupling constant, black hole charge $q$, the multipole number $l$ and parity of the perturbational field. The dependence of quasinormal frequencies on the black hole charge $q$ is also modified due to the coupling between vector field perturbation and Einstein tensor. Moreover, we also find that the vector field perturbation with even parity always decays for arbitrary multipole number $l$ and coupling constant $\alpha$. However, the vector field perturbation with odd parity grows with exponential rate  if the coupling constant $\alpha$ is smaller than certain a negative critical value $\alpha_c$, which can be fitted best by the function $\alpha_c\simeq a\sqrt{\frac{l}{l+1}}+b$  with two numerical constants ( with dimensions of length-squared) $a, b$ depended on the charge $q$. The absolute value of $\alpha_c$ decreases with the multipole number $l$ and the charge $q$, which means that in the case with the higher $l$ and the larger $q$ the unstable mode appears more easily and the growth rate becomes more stronger.

We also make a comparison among the dynamical properties of perturbational fields coupling to gravitational tensors in some black hole spacetimes.
For the scalar field coupling to Einstein tensor in the Reissner-Nordstr\"{o}m  black hole spacetime, we find that the partial wave $l=0$ always decays for arbitrary coupling constant, which is similar to that of the even parity perturbational field coupling to Einstein tensor or Weyl tensor. The nonzero-$l$ partial waves of the coupled scalar field share the similar properties of the odd parity  perturbational field coupling to Einstein tensor or Weyl tensor in the black hole spacetime.
These perturbations grow with exponential rate if the coupling strength is stronger than the certain a critical value. Moreover, we find also that the critical  coupling constant and the change of quasinormal frequencies with the coupling constant depend on the types of the coupled gravitational tensors, the black hole parameters, the multipole number $l$ and spin $s$ of perturbational fields. Finally, our results show that the constraint on the coupling constant $\alpha$ depends on the parity of vector field perturbation as it couples to Einstein tensor or Weyl tensor.
It would be of interest to generalize our study to other black hole spacetimes, such as rotating black holes etc. Work in this direction will be reported in the future.

\section{\bf Acknowledgments}

We would like to thank the referee for their useful comments and suggestions. This work was  partially supported by the National Natural Science
Foundation of China under Grant No.11275065, and the
construct program of the National Key Discipline. J. Jing's
work was partially supported by the National Natural Science
Foundation of China under Grant No. 11175065.

\appendix
\section{the wave dynamic equation for the perturbation coupling to Einstein tensor}
Inserting
the vector potential (\ref{Au}) with the separating variables (\ref{Au1}) into the modified equation of motion (\ref{WE}), we can get three independent coupled differential equations
\begin{eqnarray}
&&\mathcal{H}_0(r)\equiv r^2f[r^2-4\alpha(rf'+f-1)][ij^{lm}(r)''-\omega h^{lm}(r)']+2r^2f[r^2-2\alpha(rf''-2f)][ij^{lm}(r)'-\omega h^{lm}(r)]\nonumber\\&&+l(l+1)\omega k^{lm}(r)[r^2-\alpha(r^2f''+4rf'+2f-2)]=0,\label{Aa1}\\
&&\mathcal{H}_1(r)\equiv r^2\omega[r^2-4\alpha(rf'+f-1)][-ij^{lm}(r)'+\omega h^{lm}(r)]-l(l+1)f[r^2-\alpha(r^2f''+4rf'+2f-2)]\times\nonumber
\\&&[h^{lm}(r)-k^{lm}(r)']=0,\label{Aa2}\\
&&iP_{lm}(\cos\theta)\mathcal{H}_2(r)-
\sin^2\theta\frac{dP_{lm}(\cos\theta)}{d\cos\theta}\mathcal{H}_3(r)
=0,\label{Aa3}
\end{eqnarray}
with
\begin{eqnarray}
\mathcal{H}_2(r)&\equiv&r[r^2+\alpha(r^2f''+4rf'+2f-2)][f^2a^{lm}(r)''
+\omega^2a^{lm}(r)]
+f[rf'(r^2+\alpha(r^2f''+4rf'-2))\nonumber
\\&&+\alpha f(r^3f^{(3)}+4r^2f''-4f+4)]a^{lm}(r)'
-fl(l+1)[r+2\alpha(rf''+2f')]a^{lm}(r),
\nonumber
\\ \mathcal{H}_3(r)&\equiv&
[r+\alpha(r^2f''+4rf'+2f-2)][rf^2k^{lm}(r)''+r^2\omega^2k^{lm}(r)-ir\omega j^{lm}(r)-rf^2h^{lm}(r)']\nonumber
\\&&+f[rf'(r^2-\alpha(r^2f''+4rf'-2))-\alpha f(r^3f^{(3)}+4r^2f''-4f+4)][k^{lm}(r)-h^{lm}(r)],\nonumber
\end{eqnarray}
where $P_{lm}(\cos\theta)$ is the associated Legendre function. Obviously, Eqs.(\ref{Aa1}) and (\ref{Aa2}) are irrelevant to the function $a^{lm}(r)$. Moreover, one can find that the polynomials $\mathcal{H}_3(r)$ in Eq.(\ref{Aa3}) is related to $\mathcal{H}_0(r)$ and $\mathcal{H}_1(r)$ by
\begin{eqnarray}
l(l+1)\mathcal{H}_3(r)=rf\frac{d\mathcal{H}_1(r)}{dr}-2f(r)\mathcal{H}_1(r)
+r\omega\mathcal{H}_0(r).
\end{eqnarray}
This means that
when Eqs.(\ref{Aa1}) and (\ref{Aa2}) are satisfied, i.e. $\mathcal{H}_0(r)=0$ and $\mathcal{H}_1(r)=0$,  the equation (\ref{Aa3}) reduces naturally to $\mathcal{H}_2(r)=0$, which yields
\begin{eqnarray}
&&\frac{d^2a^{lm}(r)}{dr^2}+
\frac{d}{dr}\bigg\{\ln\frac{f[r^2+\alpha(r^2f''+4rf'+2f-2)]}{r^2}\bigg\}
\frac{da^{lm}(r)}{dr}
+\frac{\omega^2}{f^2}\;a^{lm}(r)\nonumber
\\&&-\frac{l(l+1)[r+2\alpha(rf''+2f')]}{
rf[r^2+\alpha(r^2f''+4rf'+2f-2)]}\;a^{lm}(r)=0.\label{Aa4}
\end{eqnarray}
Make use of Eq.(\ref{Aa1}), one can eliminate $k^{lm}(r)$ in Eq.(\ref{Aa2}). Defining the variable
\begin{eqnarray}
Q^{lm}(r)=\frac{r^2}{l(l+1)}\bigg(-i\omega
h^{lm}-\frac{dj^{lm}}{dr}\bigg),
\end{eqnarray}
we find that
\begin{eqnarray}
&&\frac{d^2Q^{lm}(r)}{dr^2}+\frac{d}{dr}
\bigg\{\ln\frac{f[r^2-4\alpha(rf'+f-1)]^2}{r^2[r^2-\alpha(r^2f''+4rf'+2f-2)]}
\bigg\}\frac{dQ^{lm}(r)}{dr}+\frac{\omega^2}{f^2}Q^{lm}(r)-\frac{l(l+1)}{
rf}\bigg[1-\nonumber
\\&&
\frac{\alpha(r^2f''-2f+2)}{r^2-4\alpha(rf'+f-1)}\bigg]Q^{lm}(r)+\frac{(\alpha h_3+\alpha^2h_4)Q^{lm}(r)}{r^2 f[r^2-4\alpha(rf'+f-1)][r^2-\alpha(r^2f''+4rf'+2f(r)-2)] }=0,\label{Aa6}
\end{eqnarray}
with
\begin{eqnarray}
&&h_3=-4r^2[rf'(r^2f''+2)+f(r^3f^{(3)}-r^2f''-4rf'-6)+6f^2],\nonumber\\
&&h_4=4[rf'(r^2f''+2)(r^2f''+4 rf'-2)+4 f^2(r^3f^{(3)}+4r^2f''+3r f'-2)+
\nonumber\\&&+4rff'(r^3 f^{(3)}-r^2 f''-2)-f(4r^3 f^{(3)}+5r^4 f''^2+16r^2f''+16 r^2f'^2-4-4f^3)].\nonumber
\end{eqnarray}
Therefore, both of the radial equations for the odd parity and even parity perturbation have the same form as
 \begin{eqnarray}
\frac{d^2P(r)}{dr^2}+\frac{B(r)'}{B(r)}\frac{dP(r)}{dr}
+\bigg[\frac{\omega^2}{f^2}+U(r)\bigg]P(r)=0.\label{Aa7}
\end{eqnarray}
Defining $P(r)=\Psi(r)Y(r)$ and making use of $dr_*=\frac{dr}{f}$,
we can find that Eq.(\ref{Aa7}) can be rewritten as
 \begin{eqnarray}
\frac{d^2\Psi(r)}{dr_*^2}
+\bigg[\omega^2-V(r)\bigg]\Psi(r)=0,
\end{eqnarray}
with
\begin{eqnarray}
&&Y(r)=\sqrt{\frac{f}{B(r)}},\nonumber\\
&&V(r)=-f^2\bigg[\frac{Y(r)''}{Y(r)}+\frac{B(r)'}{B(r)}\frac{Y(r)'}{Y(r)}+U(r)\bigg].
\end{eqnarray}
Comparing with Eqs.(\ref{Aa4}),Eqs.(\ref{Aa6}) and Eqs.(\ref{Aa7}), and inserting the corresponding $B(r)$ and $Y(r)$, we can get the differential equation (\ref{radial}) with their effective potentials for the odd parity and even parity perturbations, respectively.

\end{document}